\documentclass[a4paper,fleqn,usenatbib]{mnras}
\usepackage{newtxtext,newtxmath}
\usepackage[T1]{fontenc}
\usepackage{ae,aecompl}
\usepackage{multicol}
\usepackage{graphicx}	
\usepackage{amsmath}	
\usepackage{cleveref}
\usepackage{pdflscape}

\newcommand{\kms}{{\mathrm{km~s^{-1}}}}

\title[Asteroseismology of the $\delta$ Sct star BP Pegasi]{Asteroseismology of the
double-radial mode $\delta$ Scuti star BP Pegasi}

\author[Daszy\'nska-Daszkiewicz et al.]{ J. Daszy\'nska-Daszkiewicz$^{1}$\thanks{E-mail:daszynska@astro.uni.wroc.pl},
P. Walczak$^{1}$, A. A. Pamyatnykh$^{2}$, W. Szewczuk$^{1}$\\
$^{1}$Instytut Astronomiczny, Uniwersytet Wroc{\l}awski, Kopernika 11, 51-622 Wroc{\l}aw, Poland\\
$^{2}$Nicolaus Copernicus Astronomical Center, Polish Academy of Sciences, Bartycka 18, 00-716 Warsaw, Poland\\
}

\date{Accepted XXX. Received YYY; in original form ZZZ}

\pubyear{2021}

\begin{document}
\label{firstpage}
\pagerange{\pageref{firstpage}--\pageref{lastpage}}
\maketitle

\begin{abstract}
Using the ASAS data, we determine the pulsational frequencies of the high-amplitude $\delta$ Sct star BP Pegasi.
The analysis revealed only the two known, independent frequencies that we use to perform the
seismic analysis of the star. On the basis of multicoloutr Str\"omgren photometry, we independently find
that both frequencies can only be associated with radial modes which, according to the frequency ratio, 
are fundamental and first overtone modes.
The models fitting the two frequencies depend strongly on the opacity data. For low values of the mixing length parameter $\alpha_{\rm MLT}\approx 0.5$, only the OPAL seismic models in the post-main sequence phase of evolution are caught within the observed error box. Seismic models computed with the OP and OPLIB data are  much less luminous and cooler.  
They  can only reach the error box  if we increase the convection efficiency to at least $\alpha_{\rm MLT}= 2.0$.
Then, including the non-adiabatic parameter $f$ into our seismic modelling, we constrain the value of $\alpha_{\rm MLT}$.
Computing an extensive grid of seismic OPAL models and employing Monte Carlo–based Bayesian analysis, we obtain  
constraints on various parameters. In particular, the efficiency of  envelope convection can be parametrized by
 $\alpha_{\rm MLT}\in (0.5,~1.0)$, depending on the atmospheric  microturbulent velocity that amounts to $\xi_t=4$ or 8\,$\kms$.
\end{abstract}



\section{Introduction}
High-amplitude $\delta$ Scuti (HADS) stars are pulsating variables with the $V$-band range above 0.3\,mag
and constitute in some aspects a subclass of the $\delta$ Sct pulsators.
They are usually in an advanced phase of main-sequence evolution or already in a post-main sequence phase 
\citep[e.g.,][]{Breger2000}.
The HADS stars pulsate in just one or two frequencies which are assigned to radial modes.
Although, radial pulsations for the  HADS stars are most probable because of high amplitudes and the period ratio in case of double-mode
pulsators, there is some chance that non-radial modes can be present as well. Therefore, the mode identification should be confirmed by independent observables from photometric and/or spectroscopic time-series observations.
Unfortunately, a few such efforts can be found in the literature. For example, \citet{Ulusoy2013} used the $UBVRI$ time-series photometry  for the high-amplitude $\delta$ Scuti star V2367\,Cyg  but they did not get a unique identification of the mode degree $\ell$ for any of three detected frequencies. Likewise, \citet{Casas2006} did not obtain unambiguous determination of $\ell$ from multicolour diagnostic diagrams  for two frequencies of RV\,Ari.
Recently, \citet{JDD2020} applied the method of simultaneous determination of $\ell$ and the nonadiabatic parameters $f$
for the two frequencies of the prototype SX\,Phoenicis. They  used the amplitudes and phases in the Str\"omgren photometry
and successfully identified the two modes as radial ones. 

The detection of two radial modes in any star imposes very strong constraints on its mean parameters
and global chemical composition.
Despite this, there are not many papers that present  detailed seismic modelling of the HADS stars.
\citet{Petersen1996}  analyzed  pulsational models for a few HADS stars
and concluded that these are post-main sequence objects which explains their high amplitudes.
On the other hand \citet{Breger2000} suggested that high amplitudes  are related to the slow rotation of these stars,
which is  typically $V\sin i < 40~\kms$.  \citet{Casas2006} presented a more advanced modelling for RV\,Ari
including the effects of non-adiabaticity and rotation. Some tentative seismic studies were published for V2367\,Cyg
by \citet{Balona2012}.
More recently, \citet{Niu2017} constructed seismic models of the HADS star AE UMa, using the two radial-mode frequencies
as well as the period changes of the dominant mode. They obtained that AE UMa is in the post-main sequence stage of evolution.

\citet{Xue2018} made an attempt to construct seismic models for the double-mode HADS star VX Hyd
using MESA evolutionary models and the adiabatic pulsation code ADIPLS \citep{Christensen-Dalsgaard2008}.
They concluded that only post-main sequence models are suitable for this star.
Thus, the seismic modelling of $\delta$ Scuti stars pulsating in two radial modes should be definitely intensified
and the number of studied stars should be increased.

In our previous paper we presented the complex seismic modelling of the prototype SX Phoenicis \citep{JDD2020}.
This analysis consisted of a simultaneous fitting of the two radial-mode frequencies and the corresponding
values of the bolometric flux amplitude (the parameter $f$) . The effects of various model parameters were investigated.
In particular,  we showed that seismic models of SX Phe are strongly affected by the choice of opacity data.
Our extensive seismic modelling indicated a preference for OPAL tables and  the post-main sequence phase of evolution.
Besides the important constraints on the efficiency of convection in the  envelope of SX Phe,
described by the mixing length parameter $\alpha_{\rm MLT}<0.7$, 
we also obtained constraints on the microturbulent velocity in the atmosphere $\xi_t\in(4,~8)~\kms$.

Here, we present results for the double-mode HADS star BP Pegasi applying a similar approach as in the case of SX Phe.
In Sect.\,2 we give the information on BP Peg. Sect.\,3 contains the analysis of the ASAS and Str\"omgren $uvby$ data as well as the frequency determination. Mode identification based on the multicolour photometry for the two frequencies is presented in Sect.\,4.
In Sect.\, 5, we give the results of seismic modelling of BP Peg using the three sources of the opacity data and
 preliminary constraints on the efficiency of  envelope convection.
Finally, in Sect.\,6, we present an extensive seismic modelling with Monte Carlo-based Bayesian analysis.
The last section summarizes all our results.

\section{BP Pegasi}
BP Pegasi is a variable star with the average visual brightness of $V=12.05$\,mag  and the Gaia DR2 parallax of $\pi=0.8848(361)$\,mas. We discarded the Gaia EDR3 parallax $\pi=0.7959(300)$\,mas because  a quality indicator, the so-called RUWE (the renormalised unit weight error) amounts to 2.1 that is much larger than 1.0.
The RUWE  significantly greater than 1.0 (usually the limit 1.4 is adopted) could indicate that the source is non-single or otherwise problematic for the astrometric solution \citep{Lindegren2021}.

The spectral type of BP Peg in the Catalogue of SIMBAD is A0, whereas  in a more detailed study by  \citet{Rodriguez1994} there is given a range A5-F0.
The effective temperature of BP Peg was determined by several authors and its values are as follow:
$T_{\rm eff}\approx 7500$\,K \citep{Andreasen1983}, $T_{\rm eff}\in(7130,~8050)$\,K \citep{Kim1989},
$T_{\rm eff}\in(7190,~8100)$\,K  \citep{Rodriguez1992} and $T_{\rm eff}\in(6860,~8000)$\, \citep{Pena1999}. For our studies, we adopted the whole range (6860,\,8100)\,K that converts into $\log T_{\rm eff}=3.8724(361)$.
The atmospheric metallicity is  [m/H]$=-0.02$ according to \citet{McNamara1997},
[m/H]$=-0.08\pm 0.05$  according to \citet{Kim1989} whereas \citet{Rodriguez1992} estimated
the value of  [m/H]$=+0.2$. These estimates are based on the $\delta m_1$ index.

Using the Gaia DR2 parallax,  the extinction $A_V=0.216\pm 0.012$ from  the Bayestar\,2019 reddening map \citep{Green2019}  
and the bolometric correction from \citet{Flower1996}, we arrived at the luminosity range (1.137,\,1.361) that is  $\log L/{\rm L}_{\sun}=1.249(112)$. 
Using the bolometric correction from Kurucz models and taking into account different values of the atmospheric metallicity [m/H] and microturbulent velocity $\xi_t$, we derived the luminosity  range (1.127,\,1.367) that 
is $\log L/{\rm L}_{\sun}=1.247(120)$.  This value  includes determinations for [m/H]$=-0.1, 0.0,~0.2,~0.3$ 
and $\xi_t=2,~4~\kms$ and the error in the bolometric correction of about 0.015.
For the further analysis,  we adopted $\log L/{\rm L}_{\sun}=1.247(120)$ because it takes into account all possible uncertainties and includes the range derived with the Flower bolometric correction.

The analysis of spectra by \citet{Kim1989} led them to conclude that the star shows very sharp spectral lines from which they derived an upper limit  of the projected 
rotational velocity, $V\sin i$, of $18~\kms$.
Thus, BP Pegasi is a slow rotator if the inclination angle is not far from $i=90^{\circ}$.

In Fig.\,1, we showed the Hertzsprung-Russell diagram with the error box of BP Peg in a comparison with the evolutionary tracks computed for masses $M=1.8,1.9,2.0,2.1\,{\rm M}_{\sun}$, using the OPAL opacity tables \citep{Iglesias1996}, the OPAL2005 equation of state  \citep{Rogers1996, Rogers2002} and the solar mixture from \citet{Asplund2009}, hereafter  AGSS09.
We adopted the metallicity $Z=0.020$, the initial hydrogen abundance $X_0=0.70$ and the zero value of the parameter  describing overshooting from the convective core $\alpha_{\rm ov}=0.0$. Moreover, for the model with a mass $M=1.9\,{\rm M}_{\sun}$, we depicted also the tracks with $Z=0.028$, $X_0=0.73$ and $\alpha_{\rm ov}=0.2$. The evolution were computed with the Warsaw-New Jersey code \citep[e.g.,][]{Pamyatnykh1998, Pamyatnykh1999}, which takes into account the mean effect of the centrifugal force,
assuming a solid-body rotation and constant global angular momentum during evolution.
The convective transport in the envelope is treated in the framework of the standard mixing-length theory.
The tracks in Fig.\,1 were computed with the mixing length parameter $\alpha_{\rm MLT}=0.5$ and the initial velocity of rotation $V_{\rm rot,0}=25~\kms$.

\begin{figure}
\includegraphics[width=\columnwidth,clip]{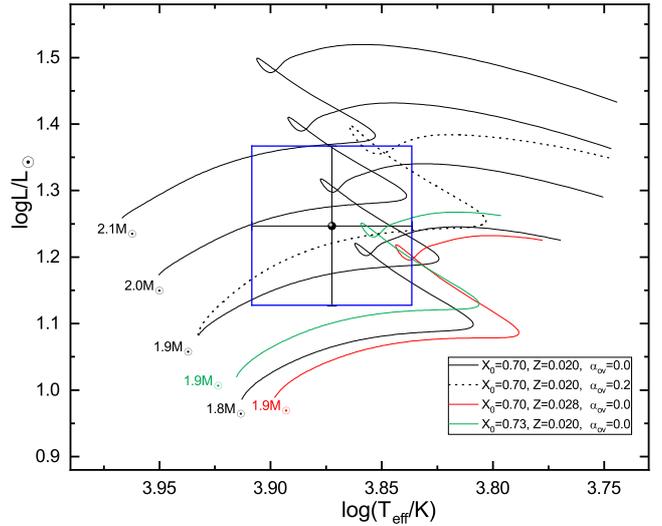}
\caption{The HR diagram with the position of BP Pegasi. The range of the effective temperature includes all values
found in the literature. The luminosity was derived adopting the Gaia DR2 parallax. The evolutionary tracks, for masses $M=1.8,1.9,2.0,2.1\,{\rm M}_{\sun}$, were computed with the OPAL opacity tables and the AGSS09 solar mixture. There is shown the effect of the initial hydrogen abundance $X_0$, metallicity $Z$ and the convective overshooting parameter $\alpha_{\rm ov}$. The assumed values of $X_0$, $Z$ and $\alpha_{\rm ov}$ are given in the legend. The mixing length parameter was $\alpha_{\rm MLT}=0.5$ and the initial velocity of rotation $V_{\rm rot,0}=25~\kms$.}
\label{fig1}
\end{figure}

The variability of BP Pegasi was firstly reported by \citet{Masani1954}. Then, \citet{Broglia1959} showed that it pulsates in two modes with a dominant  period of
0.10954347\,d. The range of a brightness changes was about 0.45\,mag.  Moreover, \citet{Broglia1959} found a modulation period on the order of 0.37\,d.  Based on the visual observations Figer (1983) derived the values of these periods at $P_1= 0.109543375$\,d and $P_2 = 0.084510$\,d.
The $uvby\beta$ photometry and the first time series spectroscopy was gathered by \citet{Kim1989}.
They found the total range of the radial velocity variations of $36~\kms$.
Detailed frequency analysis was done by   \citet{Rodriguez1992} who found nine frequencies
in the $V$ photometry of \citet{Broglia1959} but only two of them were independent.
\citet{Rodriguez1992}  considered BP Peg as a classical large amplitude $\delta$ Scuti star and
gained also the new photometry in the Str\"omgren $uvby\beta$ filters
and determined the amplitudes and phases for the two independent frequencies.
Another  $uvby\beta$ photometric study was  carried out by \citet{Pena1999}  who suggested also that  the dominant frequency
of BP Peg is the radial fundamental mode from the pulsational constant.
Although some authors have attempted to assign BP Peg to the RR Lyr type \citep{Figer1983} and it is assigned to that type in the Catalogue of SIMBAD,  BP Peg is classified  in the General Catalog of Variable Stars (GCVS, 2020)
as a high amplitude $\delta$ Scuti star (HADS).

The observed period ratio of 0.7715 indicates that BP Peg pulsates, most likely, in two radial modes, fundamental and first overtone, as already suggested by \citet{Fitch1976}.  Besides, based on the period ratio, \citet{Cox1979} concluded that the star is a normal Population I $\delta$ Scuti star.
 An attempt to make an independent identification of the mode degree $\ell$ of the two frequencies has been undertaken by \citet{Balona1999}  on the basis of the photometric amplitudes and phases. However, the authors did not get a unique discrimination between the radial and dipole mode.

\section{Data Analysis}


As we have mentioned in the Introduction  \citet{Rodriguez1992} derived  nine frequencies from the
\citet{Broglia1959} data. We listed them in Table\,\ref{tab:Broglia_freq_or}. The data of \citet[][]{Broglia1959} consisted of 861 observational points and spanned 127 days which corresponds to the Rayleigh resolution of $1/T=0.008\,\mathrm d^{-1}$. BP Peg was also observed in the framework of the All-Sky Automated Survey 
\citep[ASAS,][]{Pojmanski2002}, specifically in the phase ASAS-3.

\begin{table}
	\centering
	\caption{Frequencies, amplitudes and phases found by \citet{Rodriguez1992} in the \citet{Broglia1959} observations of BP Peg. }
\label{tab:Broglia_freq_or}
	\begin{tabular}{r r r r } 
		\hline
\hline
      ID       &  Frequency         &    $A_\mathrm{V}$ & $\phi_\mathrm{V}$    \\
               & $[\mathrm d^{-1}]$           &      [mmag]       &   [rad] \\
\hline
      $\nu_1$  &    9.1291        & 190     &  1.929\\
               &                  & $\pm 1$ & $\pm 0.006$\\
     $2\nu_1$  & 18.2582          & 60      & 0.07 \\
               &                  & $\pm 1$ & $\pm 0.02$ \\
    $3\nu_1$   & 27.3873          & 19      & 4.48\\
               &                  & $\pm 1$ & $\pm 0.06$ \\
     $\nu_2$   & 11.8329          & 16      & 1.48\\
               &                  & $\pm 1$ & $\pm 0.07$ \\
$\nu_1+\nu_2$  & 20.9620          & 14      & 5.98\\
               &                  & $\pm 1$ & $\pm 0.08$ \\
    $4\nu_1$   & 36.5164          & 8       & 2.9\\
               &                  & $\pm 1$ & $\pm 0.1$ \\
$\nu_2-\nu_1$  & 2.7038           &  7      & 3.9\\
               &                  & $\pm 1$ & $\pm 0.1$ \\
$2\nu_1+\nu_2$ & 30.0911          & 6       & 4.2\\
               &                  & $\pm 1$ & $\pm 0.2$ \\
$3\nu_1+\nu_2$ & 39.2202          & 5       & 2.2\\
               &                  & $\pm 1$ & $\pm 0.2$ \\
\hline
	\end{tabular}
\end{table}

\begin{figure*}
	\includegraphics[angle=270, width=2\columnwidth]{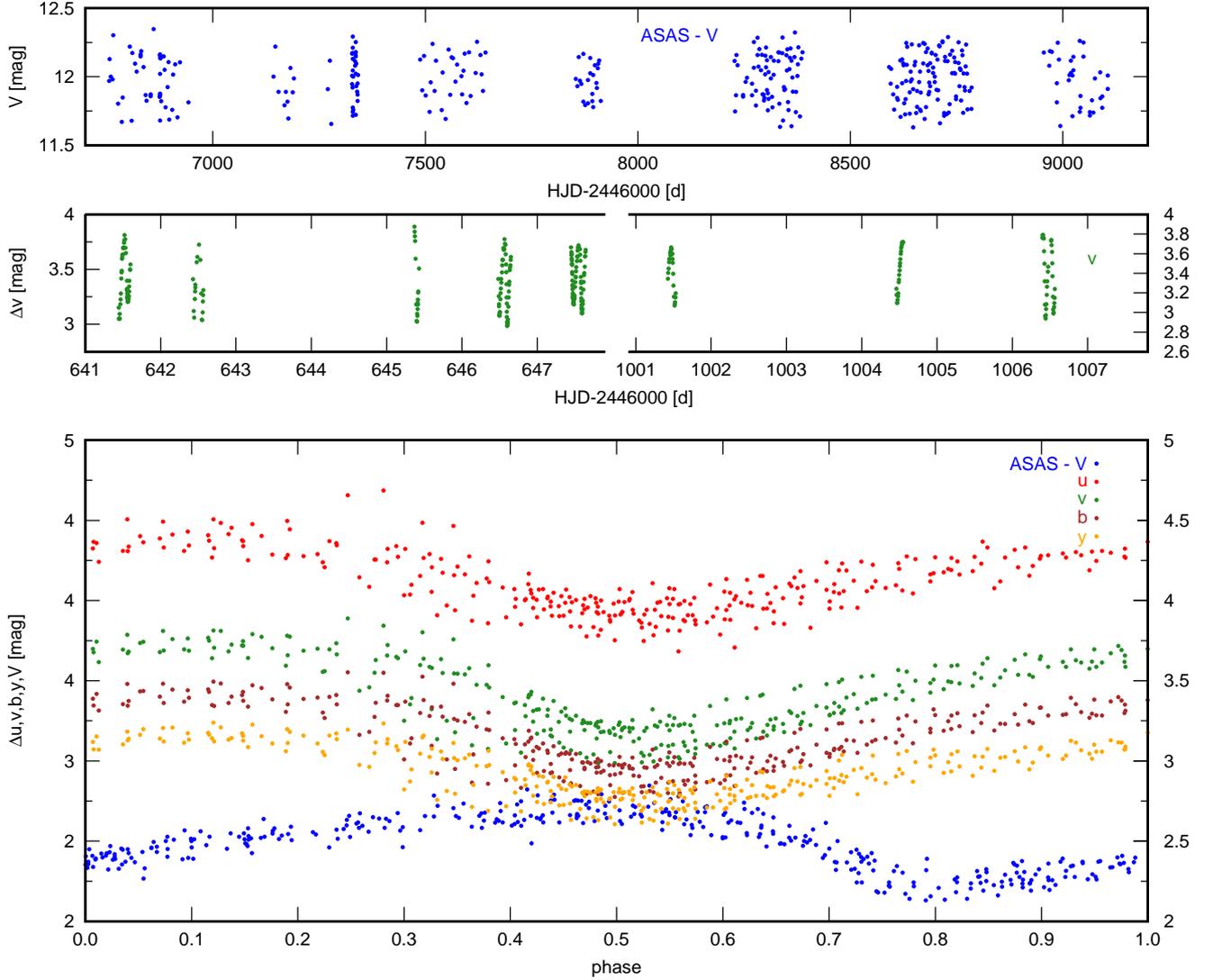}
    \caption{Photometric observations of BP Peg. The two top panels show the ASAS3-V and Str\"omgren $v$ data from Rodriguez.
    The bottom panel contains the ASAS and Str\"omgren $uvby$ observations phased with $\nu_1= 9.128797$\,d$^{-1}$. }
    \label{fig:freq_analysis}
\end{figure*}

\begin{figure*}
	\includegraphics[angle=270, width=2\columnwidth]{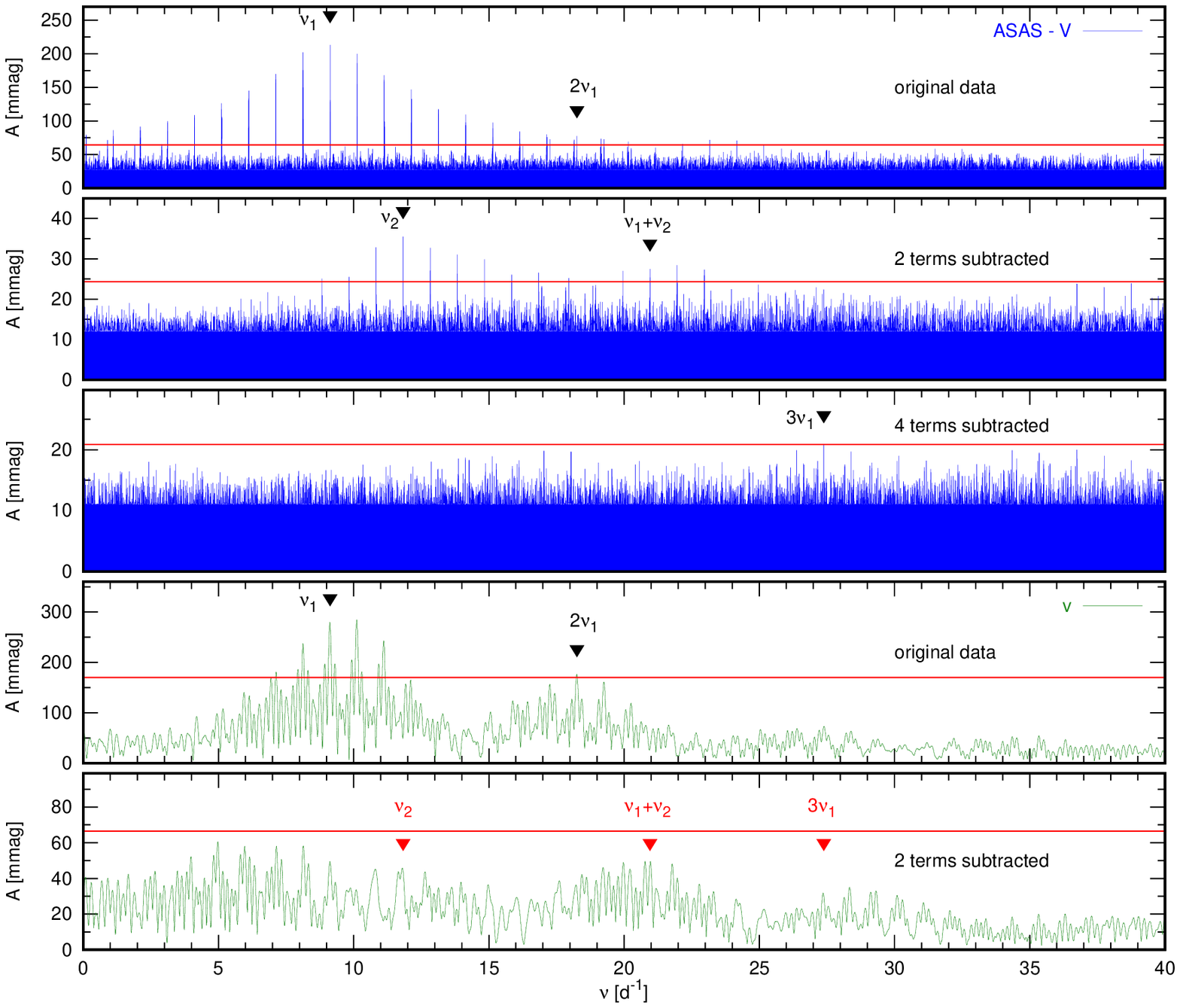}
    \caption{Periodograms from photometric observations of BP Peg. Panels from top to bottom show:
     periodograms for the ASAS-3 data (on the original data ,  after a subtraction of 2 terms and after a subtraction of 4 terms),
    periodograms for the $v$ data (on the original data, after a subtraction of 2 terms). Black triangles mark significant frequencies. Red triangles mark known frequencies but insignificant in the given periodogram. Red horizontal lines indicate the $4S/N$ level.}
    \label{fig:freq_analysis_2}
\end{figure*}

Here we analyzed both the Str\"omgren photometry \citep{Rodriguez1992} as well as the ASAS-3 V-band photometry.
We start from the ASAS data that span 2350 days (see Fig.\,\ref{fig:freq_analysis}, top panel)
which gives the Rayleigh resolution of $0.0004\,\mathrm d^{-1}$.
The ASAS-3 photometry was constructed with five different apertures and  each observational point
has been assigned a quality flag (the best points have the A flag, the worst - the D flag).
We started from removing points with the quality flag C and D.
Then, we calculated  amplitude periodograms by means of a discrete Fourier transform
\citep{Deeming1975, Kurtz1985} for photometric 
data from two apertures which had the smallest mean errors.
Periodograms were calculated with the resolution of $0.000004\,\mathrm d^{-1}$ up to 40 d$^{-1}$.
As a noise level we adopted the mean amplitude in the periodogram calculated for the data before subtraction of a  considered frequency.
In the periodogram for the data with the second smallest mean error the signal to noise ratio ($S/N$)
for the highest amplitude peak was slightly higher. Therefore, we chose this data set
for further analysis. However, we note that frequencies from data from both apertures agree within the errors.
Moreover, in the residuals we found seven obvious outliers that were also removed from the original data.
Finally, we were left with 357 data points. For this cleaned data set we recalculated the periodogram
(see Fig.\,\ref{fig:freq_analysis_2}, top panel) and performed the standard prewhitening procedure.
Although daily aliases are high, the highest peaks in subsequent periodograms are in agreement with frequencies found by \citet{Rodriguez1992}
in \citet{Broglia1959} data (compare  Table\,\ref{tab:Broglia_freq_or} and \ref{tab:freq_my}).
As a significant frequency peaks we considered those with $S/N>4$ (Breger 1993, Kuschnig et al. 1997). Periodograms for data after subtraction two and four frequency peaks are shown
in Fig.\,\ref{fig:freq_analysis_2} (second and third panel from the top, respectively).
The frequency peak seen in the third panel from the top in Fig.\,\ref{fig:freq_analysis_2}  is slightly below our significance threshold. However, it agrees with the third harmonic of $\nu_1$. Therefore we accepted it as a real one.
Finally, we detected 5 frequencies: two independent,
first and second harmonics of $\nu_1$ and  combination $\nu_1+\nu_2$ (see Table\,\ref{tab:freq_my}).
The ASAS data phased with $\nu_1$ are shown as blue dots in the third panel from the top in Fig\,\ref{fig:freq_analysis}.
\begin{table}
	\centering
	\caption{The value of the observed frequencies, amplitudes and phases found by us in the ASAS data.}
\label{tab:freq_my}
	\begin{tabular}{r r r r r} 
\hline
      ID       &  Frequency         & $S/N$ &   $A_\mathrm{V}$ & $\phi_\mathrm{V}$ \\
               & $[\mathrm d^{-1}]$ &       &    [mmag]        &   [rad] \\
\hline
      $\nu_1$  &    9.128797        & 13.23 &      208   &  5.3      \\
               &    $\pm 0.000004$  &       &    $\pm 4$ & $\pm 0.1$ \\
      $2\nu_1$ &   18.25759         & 8.01  &  60        &  0.9       \\
               &    $\pm 0.00001$   &       &    $\pm 4$ &  $\pm 0.4$ \\
      $\nu_2$  &   11.83315         & 5.84  & 34         &  4.4       \\
               &    0.00003         &       & $\pm 4$    & $\pm 0.6$  \\
$\nu_1+\nu_2$  &   20.96188         & 4.66  &  26        &   2        \\
               &  $\pm 0.00004$     &       & $\pm 4$    &   $\pm 1$  \\
     $3\nu_1$  &   27.38633         & 3.99  &  23        &  4         \\
               &   $\pm 0.00004$    &       & $\pm 4$    &  $\pm 1$   \\
\hline
	\end{tabular}
\end{table}

In the next step, we reanalysed Str\"omgren photometry  of \citet{Rodriguez1992}.
These data consist of 299 observations in each filter that span 365 days that gives the Rayleigh resolution of 0.003\,d$^{-1}$.
Observations in  the $v$ filter are shown in the second panel from the top in Fig.\,\ref{fig:freq_analysis}.
In each Str\"omgren filter we performed the same periodogram analysis as in the case of ASAS data.
However, this time in all four filters we found the highest peak at a frequency of about 10.115\,d$^{-1}$, i.e., at the position of a daily alias of $\nu_1$. 
As an example we depict the periodogram in the Str\"omgren $v$ filter  (the second panel from the bottom in Fig.\,\ref{fig:freq_analysis_2}).
In the case of the $u$ data only this frequency (or its aliases) satisfy our significance criterion.
In the $v$ and $y$ filters we found
the second significant frequency at about 17.280\,d$^{-1}$ which is the alias of $2\nu_1$. 
We note that in the periodogram for the original $v$ observations the second highest peak is 2$\nu_1$.
However, in the periodogram calculated for the residuals after subtraction of $\nu_1$ the highest peak is a daily alias of 2$\nu_1$.
Finally, in the $b$ filter we found second significant frequency at 18.269\,d$^{-1}$ which is $2\nu_1$. In the bottom panel of  Fig.\,\ref{fig:freq_analysis_2}, we show the periodogram for the $v$ data after a subtraction of the two terms ($\nu_1$ and $2\nu_1$). As one can see there are maxima in the positions of known frequencies but they are below our $S/N$ significance criterion of 4.0.
Therefore, in our further analysis (see the next Section) we decided to use amplitudes and phases in the $uvby$ filters fitted by \citet{Rodriguez1992} who fixed frequencies to the values from Table\,\ref{tab:Broglia_freq_or} (their Table 12, the second column).
We list these data in Table\,\ref{tab:amp_phas_Rod}. We should mention that the differences 
in the values of frequencies $\nu_1$ and $\nu_2$ between those used by \citet{Rodriguez1992} and our determinations from ASAS are about 0.0003\,d$^{-1}$.
With such small differences the frequency ratio $\nu_1/\nu_2$ agrees up the fourth decimal place, which is actually the limit of numerical accuracy.

Observations in the Str\"omgren $uvby$ filters phased with $\nu_1$ are shown in the bottom panel of Fig.\,\ref{fig:freq_analysis}. However, we note that there is a large phase shift between the $uvby$ light curves and the ASAS light curve. This is most probably due to miscounting of time in one of the data sets.
We would also like to point out that our values of frequencies are in agreement (within the Rayleigh resolution
of $0.008\,\mathrm d^{-1}$) with those found by \citet{Rodriguez1992} in \citet{Broglia1959} data. On the other hand, the values of $2\nu_1$ in the $u$ and $b$ filters found by \citet{Rodriguez1992} in its own Str\"omgren photometry differ more than the Rayleigh resolution (0.003\,d$^{-1}$) from our ASAS frequencies.
The same is true for $\nu_2$ in the $uvy$ filters, $\nu_1+\nu_2$ in the $by$ filters as well as for $3\nu_1$
in  the $uby$ filters.
\begin{table*}
	\centering
	\caption{The photometric amplitudes and phases determined by \citet{Rodriguez1992} in the $uvby$ filters. Frequencies listed in the second column are repeated from Table\,\ref{tab:Broglia_freq_or}.}
\label{tab:amp_phas_Rod}
	\begin{tabular}{r r r r r r  r r r r} 
\hline
      ID       &  Frequency         &   $A_\mathrm{u}$ & $\phi_\mathrm{u}$&   $A_\mathrm{v}$ & $\phi_\mathrm{v}$&   $A_\mathrm{b}$ & $\phi_\mathrm{b}$&   $A_\mathrm{y}$ & $\phi_\mathrm{y}$\\
               & $[\mathrm d^{-1}]$           &    [mmag]             &   [rad] &    [mmag]             &   [rad] &    [mmag]             &   [rad] &    [mmag]             &   [rad]\\
\hline
      $\nu_1$  &    9.1291          & 231     & 3.764     &   287   & 3.732     &   241  &  3.714     & 200     & 3.694 \\
               &                    & $\pm 5$ &$\pm 0.023$& $\pm 3$ &$\pm 0.012$& $\pm 3$&$\pm 0.014$ & $\pm 3$ & $\pm 0.017$ \\
      $2\nu_1$ &   18.2582          & 74      & 3.779     &    93   & 3.638     &    85  &  3.645     & 63      & 3.708 \\
               &                    & $\pm 5$ &$\pm 0.73$ & $\pm 3$ &$\pm 0.037$& $\pm 3$&$\pm 0.040$ & $\pm 3$ & $\pm 0.053$ \\
      $\nu_2$  &   11.8329          & 56      & 4.499     &    54   & 4.387     &    47  &  4.392     & 38      & 4.229 \\
               &                    & $\pm 5$ &$\pm 0.094$& $\pm 3$ &$\pm 0.062$& $\pm 3$&$\pm 0.069$ & $\pm 3$ &  $\pm 0.083$ \\
$\nu_1+\nu_2$  &   20.9620          & 34      & 4.890     &    35   & 4.662     &    32  &  4.711     & 27      & 4.770 \\
               &                    & $\pm 5$ &$\pm 0.152$& $\pm 3$ &$\pm 0.095$& $\pm 3$&$\pm 0.100$ & $\pm 3$ &  $\pm 0.117$ \\
     $3\nu_1$  &   27.3873          & 27      & 3.918     &    32   & 3.644     &    29  &  3.642     & 25      & 3.548 \\
               &                    & $\pm 5$ &$\pm 0.189$& $\pm 3$ &$\pm 0.104$& $\pm 3$&$\pm 0.108$ & $\pm 3$ &  $\pm 0.125$ \\
\hline
	\end{tabular}
\end{table*}

\section{Identification of the mode degree $\ell$}
The period ratio of the two modes of BP Pegasi amounts to 0.77146.  As mentioned in Sect.\,2, this strongly suggests that these are two radial modes; more specifically fundamental and first overtone.
Here, we want to independently confirm this identification using photometric observables.

To make the mode identification from the photometric amplitudes and phases, we rely on the method based on a simultaneous determination of the mode degree $\ell$, the intrinsic mode amplitude $\varepsilon$ multiplied by $Y_{\ell}^m(i,0)$ and the non-adiabatic parameter $f$ for a given frequency \citep{JDD2003, JDD2005}.
The value of $\varepsilon$ defines the relative radial displacement at the surface caused by a pulsational mode with the angular frequency $\omega$. The factor $Y_{\ell}^m(i,0)$ is the spherical harmonic that depends on the inclination angle $i$. In the case of radial modes we get the $\varepsilon$ value itself, because $Y_{\ell}^m(i,0)=1$. The non-adiabatic parameter $f$ is the ratio of the relative flux variation to the relative radial displacement at the photosphere level.
Both, $\varepsilon$ and $f$ have to be regarded as complex numbers because pulsations are non-adiabatic.
The method requires models of stellar atmospheres and in this paper we use Vienna (NEMO) models \citep{Heiter2002} that include turbulent convection treatment from \citet{Canuto1996}.

The method has been applied many times and a detailed description as well as the formulae underlying it can be found in \citet{JDD2003, JDD2005}.  Therefore, we will omit mathematical details here. 
The purpose of the method is to find a degree $\ell$ for which there is a clear minimum in the difference between the calculated and observed photometric amplitudes and phases.  The goodness of the fit 
can be measured in the terms of a discriminant:
$$\chi^2=\frac1{2N-N_p} \sum_{i=1}^N  \frac{ \left|{\cal A}^{obs}_{\lambda_i} - {\cal A}^{cal}_{\lambda_i}\right|^2 }{ |\sigma_{\lambda_i}|^2},\eqno(1)$$
where ${\cal A}^{obs}$ and ${\cal A}^{cal}$ denote the complex observational and the calculated amplitudes
from the empirical values of $\varepsilon$ and $f$, respectively. $N$ is the number of passbands $\lambda$
 and $N_p$ is the number of parameters to be determined. The method yields two complex parameters, $\tilde\varepsilon$
 and $f$, thus $N_p=4$. The observational errors $\sigma_{\lambda}$ are expressed as
$$|\sigma_\lambda|^2= \sigma^2 (A_{\lambda})  +  A_{\lambda}^2 \sigma^2(\varphi_\lambda), \eqno(2)$$
where $A_{\lambda}=|{\cal A}_\lambda|$ and $\varphi_{\lambda}=arg({\cal A}_\lambda)$,
are the values of the amplitude and phase, respectively. The value of $\ell$ and associated empirical values of $\varepsilon$  and $f$  are taken as the most probable for the $\ell=\ell_1$  at which the $\chi^2(\ell_1)$ reaches a minimum. 

In Fig.\,4, we show the values of the discriminant $\chi^2$ as a function of $\ell$ for the dominant frequency $\nu_1=9.128797$\,d$^{-1}$. We considered several values of $T_{\rm eff}$ and $\log g$ to include
the entire observed error box. Besides, we checked the effect of the atmospheric parameters, [m/H] and $\xi_t$. As one can see for all values of ($\log T_{\rm eff},~\log g$) and all considered pairs of ([m/H],\,$\xi_t$) the clear minimum of $\chi^2$ is at $\ell=0$. Thus, there is no doubt that the dominant mode is radial.
\begin{figure*}
	\includegraphics[width=175mm,clip]{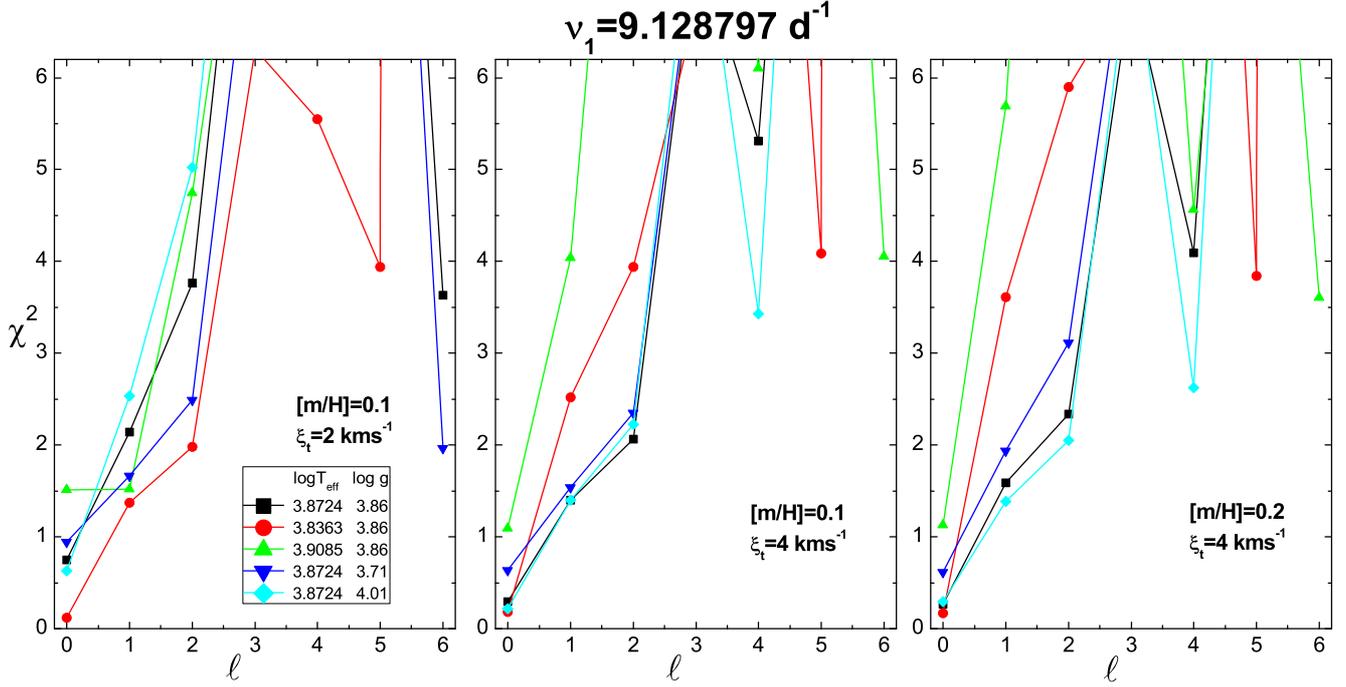}
	\caption{The discriminant $\chi^2$ as a function of $\ell$ for the dominant frequency $\nu_1$ of BP Peg
 for five combinations of the effective temperatures $\log T_{\rm eff}$ and surface gravity $\log g$, and
 three combinations of the atmospheric metallicity [m/H] and microturbulent velocity $\xi_t$.}
	\label{fig4}
\end{figure*}

The similar plots were drawn for the second frequency in Fig.\,5. Also in this case there is a strong indication
for $\ell=0$. Thus, we obtained an independent and unique identification of the two modes of BP Peg
and confirmed that they are radial as suggested by the period ratio.
\begin{figure*}
	\includegraphics[width=175mm,clip]{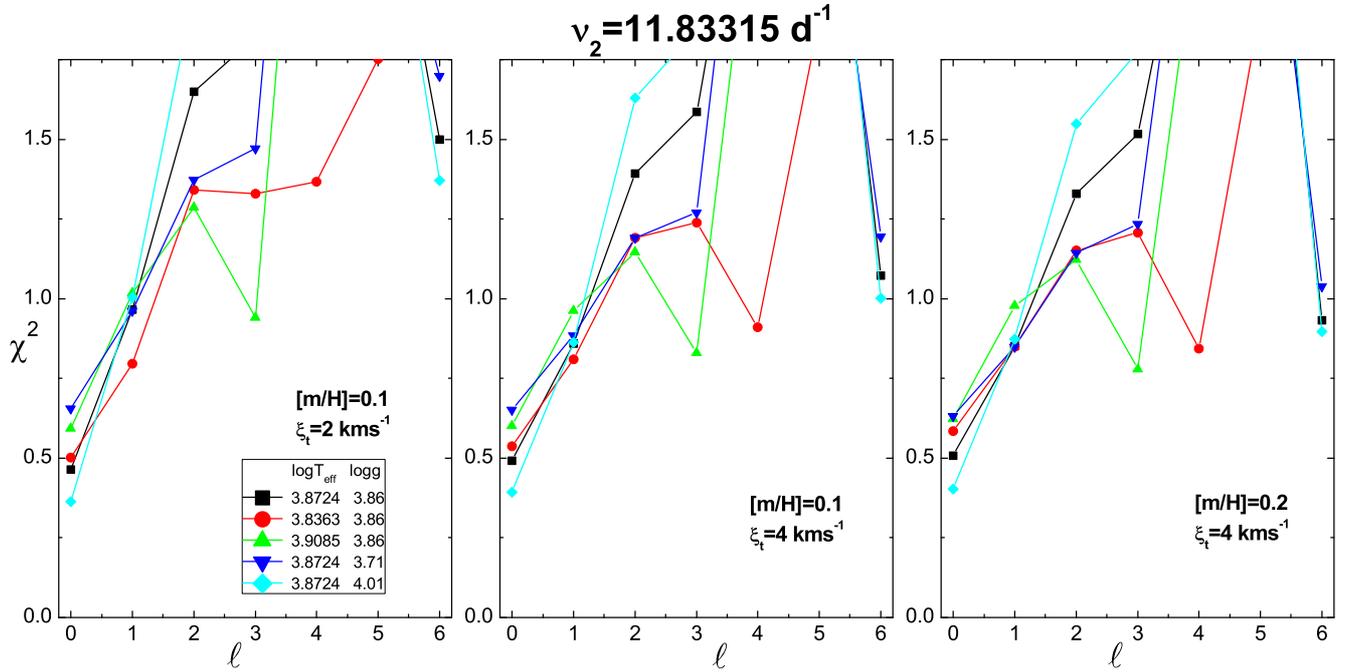}
	\caption{The same as in Fig.\,4 but for the frequency $\nu_2$.}
	\label{fig5}
\end{figure*}

Besides, the empirical values of $f$ can be compared with the theoretical counterparts from linear computations of stellar pulsations. This will be done in the next Section. In contrast, the empirical values of $\varepsilon$ cannot be compared with theoretical predictions  because the radius amplitude is not determinable in the framework of the linear theory of stellar pulsations which we use here.
However, from the empirical values of $\varepsilon$  we can estimate the amplitude
of radial velocity variations, $A(V_{\rm rad})$, due to pulsations and, in the case of radial modes, we get
an absolute value of the radius changes for a given pulsational mode.
\begin{table}
	\centering
	\caption{The empirical values of the intrinsic mode amplitude $\varepsilon$ and the resulting amplitude
of the radial velocity variations of the two pulsational modes of BP Pegasi. The examples are given for the two values of $T_{\rm eff}$, $\log g$, [m/H] and $\xi_t$.}
	\begin{tabular}{cccccrcc}
		\hline
		\multicolumn{4}{c}{$\nu_1=9.128797$ [d$^{-1}$]} \\
		\hline
$\log T_{\rm eff}$ & $\log g$ & [m/H] &    $\xi_t$   & $|\varepsilon|$ & $A(V_{\rm rad})$  \\
                     &          &       &    [$\kms$]  &               &   [$\kms$]        \\
		\hline
		\hline
	3.8724 & 3.86    &   0.2  &   2   & 0.014(4) &  12.3(3.4)  \\
	3.8363 & 3.86    &   0.2  &   2   & 0.012(2) &  10.4(1.7)  \\
	3.8724 & 4.01    &   0.2  &   2   & 0.021(4) &  15.2(3.2)  \\
		\hline
	3.8724 & 3.86    &   0.3  &   2   & 0.011(3) &   9.8(2.7)  \\
    3.8363 & 3.86    &   0.3  &   2   & 0.014(2) &  12.2(1.3)  \\
    3.8724 & 4.01    &   0.3  &   2   & 0.016(4) &  12.0(2.7)  \\
		\hline
	3.8724 & 3.86    &   0.3  &   4   & 0.012(2) &  10.3(2.1)  \\
    3.8363 & 3.86    &   0.3  &   4   & 0.021(2) &  18.5(1.5)  \\
    3.8724 & 4.01    &   0.3  &   4   & 0.016(3) &  11.5(2.2)  \\
		\hline		
		\multicolumn{4}{c}{$\nu_2=11.83315$ [d$^{-1}$]} \\
		\hline
$\log  T_{\rm eff}$& $\log g$ & [m/H] & $\xi_t$  & $|\varepsilon|$ & $A(V_{\rm rad})$ \\
                    &          &       & [$\kms$] &               &   [$\kms$]       \\
        \hline
		\hline
	3.8724 & 3.86    &   0.2  &   2   & 0.008(2) &   8.8(2.6)  \\
    3.8363 & 3.86    &   0.2  &   2   & 0.006(3) &   7.2(3.0)  \\
    3.8724 & 4.01    &   0.2  &   2   & 0.013(3) &  11.9(3.2)  \\
\hline
    3.8724 & 3.86    &   0.3  &   2   & 0.007(2) &   8.3(2.7)  \\
    3.8363 & 3.86    &   0.3  &   2   & 0.006(3) &   6.8(3.2)  \\
    3.8724 & 4.01    &   0.3  &   2   & 0.012(4) &  11.0(3.3)  \\
\hline
    3.8724 & 3.86    &   0.3  &   4   & 0.006(2) &   6.4(2.8)  \\
    3.8363 & 3.86    &   0.3  &   4   & 0.005(3) &   5.3(3.2)  \\
    3.8724 & 4.01    &   0.3  &   4   & 0.009(4) &   8.3(3.3)  \\
  \hline
	\end{tabular}
\end{table}

In Table\,4, we give the derived values of $|\varepsilon|$ and $A(V_{\rm rad})$ of the two modes
for two different values of the effective temperature, surface gravity, atmospheric metallicity and
microturbulent velocity. As one can see all these parameters have a significant effect on $|\varepsilon|$ and $A(V_{\rm rad})$.
The predicted amplitude of the relative radius changes for the dominant mode is between 1.1\% and 2.1\%,
what results in the predicted amplitude of the radial velocity variations in the range $9.8 - 18.5~\kms$.
For the second mode we obtained the radius change of (0.5\% - 1.3\%) and the radial velocity amplitude
of $5.3 - 11.9~\kms$. 
\citet{Kim1989} obtained the total range of  radial velocity variations of   $36 \pm 3~\kms$, which 
gives an amplitude from  $16.5~\kms$ to $19.5~\kms$. This observed amplitude results from pulsations in the two radial modes
and is consistent with our estimates for $\nu_1$ and $\nu_2$.

\section{Seismic modelling of BP Peg}

\subsection{Fitting the two radial modes}
Having unambiguous identification of the mode degree $\ell$, we can now construct seismic models that reproduce 
the two frequencies as the radial modes. We started from finding the models that reproduce the dominant frequency $\nu_1=9.128797$d$^{-1}$. To this aim we use the linear nonadiabatic code for stellar pulsations of Dziembowski 
\citep{Dziembowski1977a, Pamyatnykh1999}.
The code adopts the frozen convection approximation, i.e., the convective flux does not change during the pulsations. The effects of rotation on pulsational frequencies are taken into account up to the second order in the framework of perturbation theory.

We performed computations for the three opacity tables: OPAL \citep{Iglesias1996}, OP \citep{Seaton2005} and OPLIB \citep{Colgan2015, Colgan2016}.
Regardless of the period ratio of the two frequencies, all models within the error box clearly indicate that the dominant radial mode can be only fundamental. Thus, the second frequency can only be  the radial first overtone.
However, the models reproducing the frequency $\nu_1$ can be in the three evolutionary stages: main sequence (MS),
overall contraction (OC) and hydrogen-shell burning (HSB).
In Fig.\,6, we have drawn the lines of constant period corresponding to the dominant frequency of BP Peg
for the models computed with the OPAL (the left panel), OP (the middle panel) and OPLIB (the right panel) data.
The example is shown for the following parameters:
the initial hydrogen abundance $X_0=0.70$, metallicity $Z=0.020$, initial rotational velocity $V_{\rm rot,0}=
25~\kms$, the mixing length parameter $\alpha_{\rm MLT}=0.5$ and no overshooting from the convective core.

\begin{figure*}
\centering
\includegraphics[clip,width=0.33\linewidth,height=58mm]{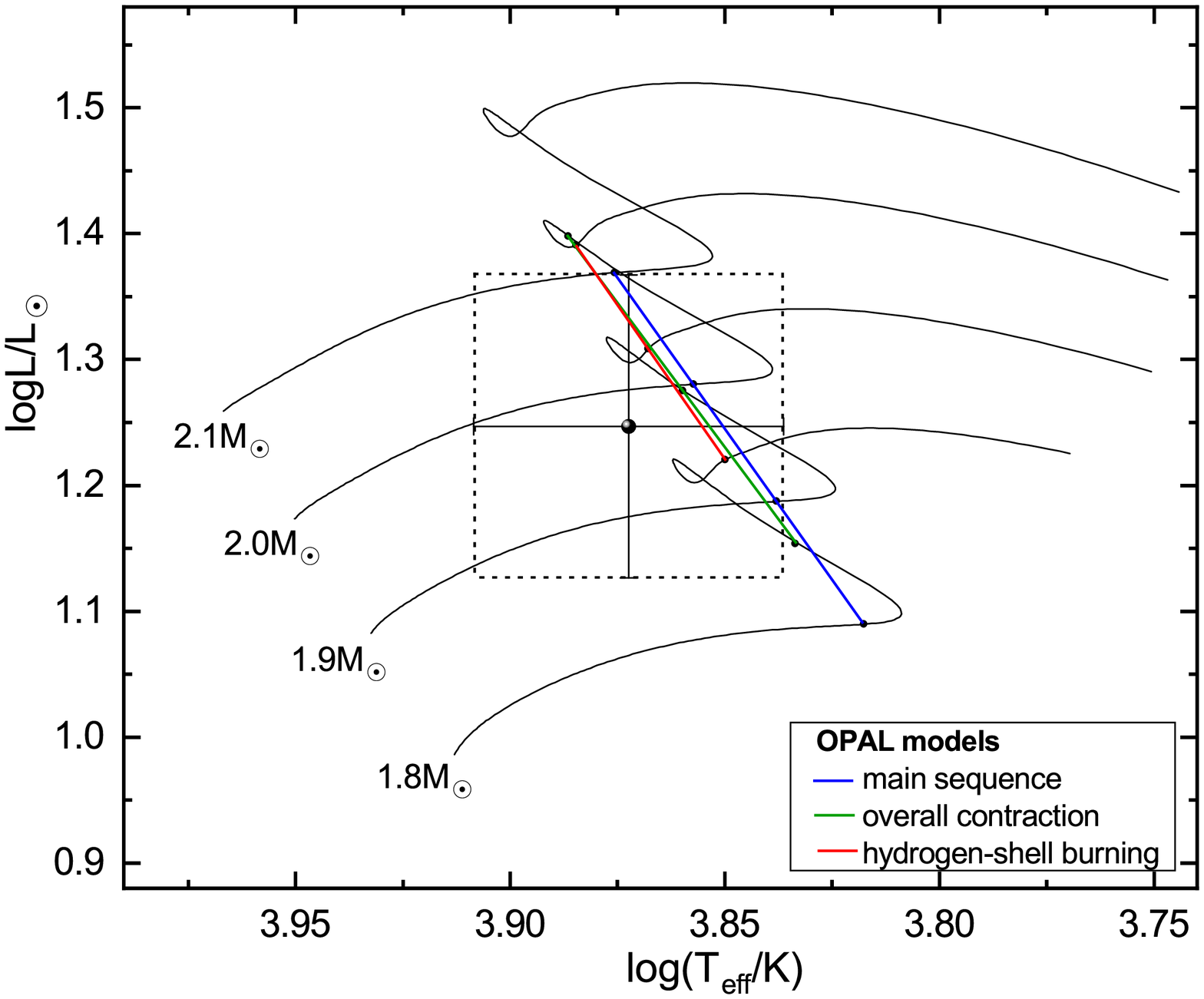}
\includegraphics[clip,width=0.33\linewidth,height=58mm]{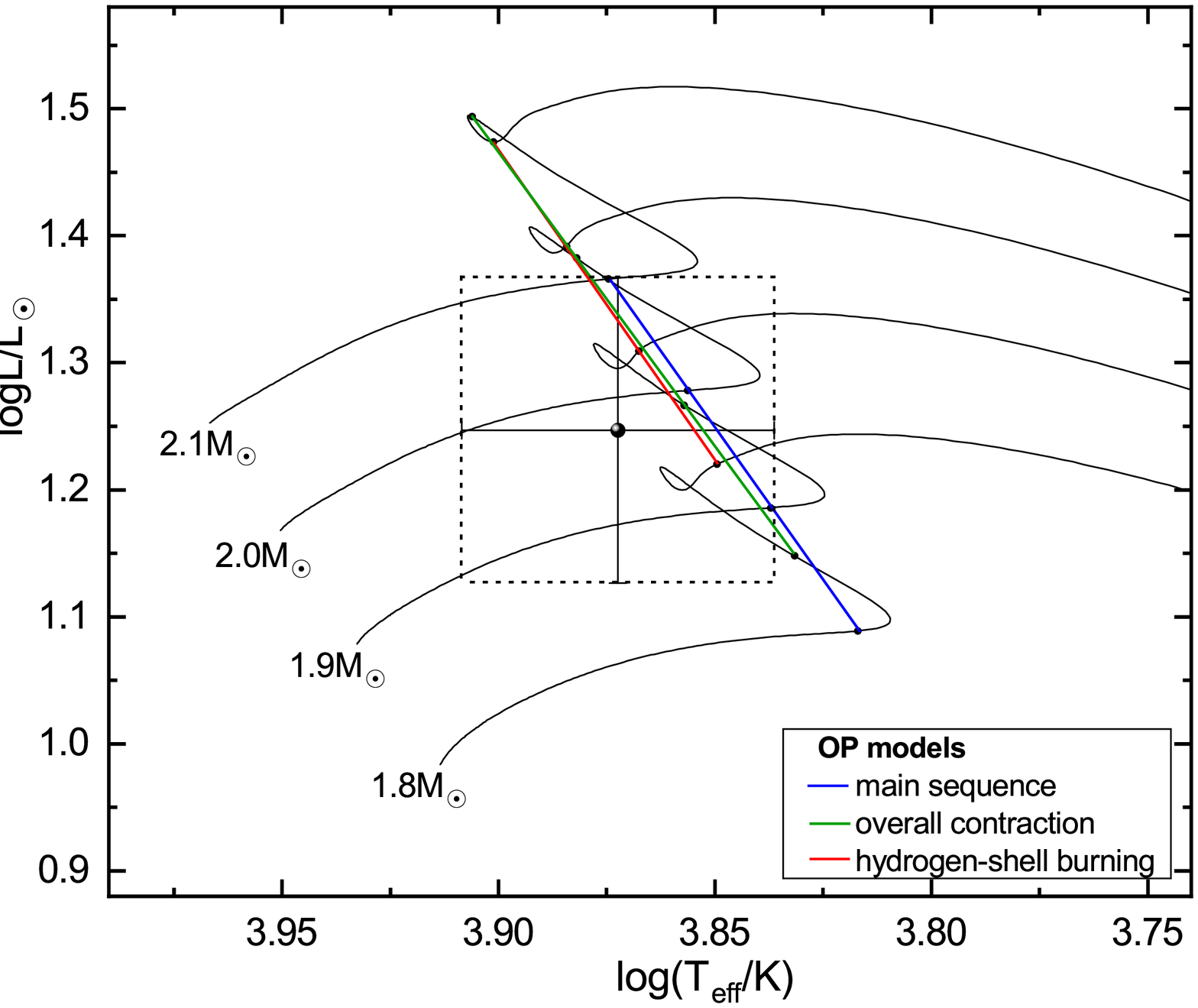}
\includegraphics[clip,width=0.33\linewidth,height=58mm]{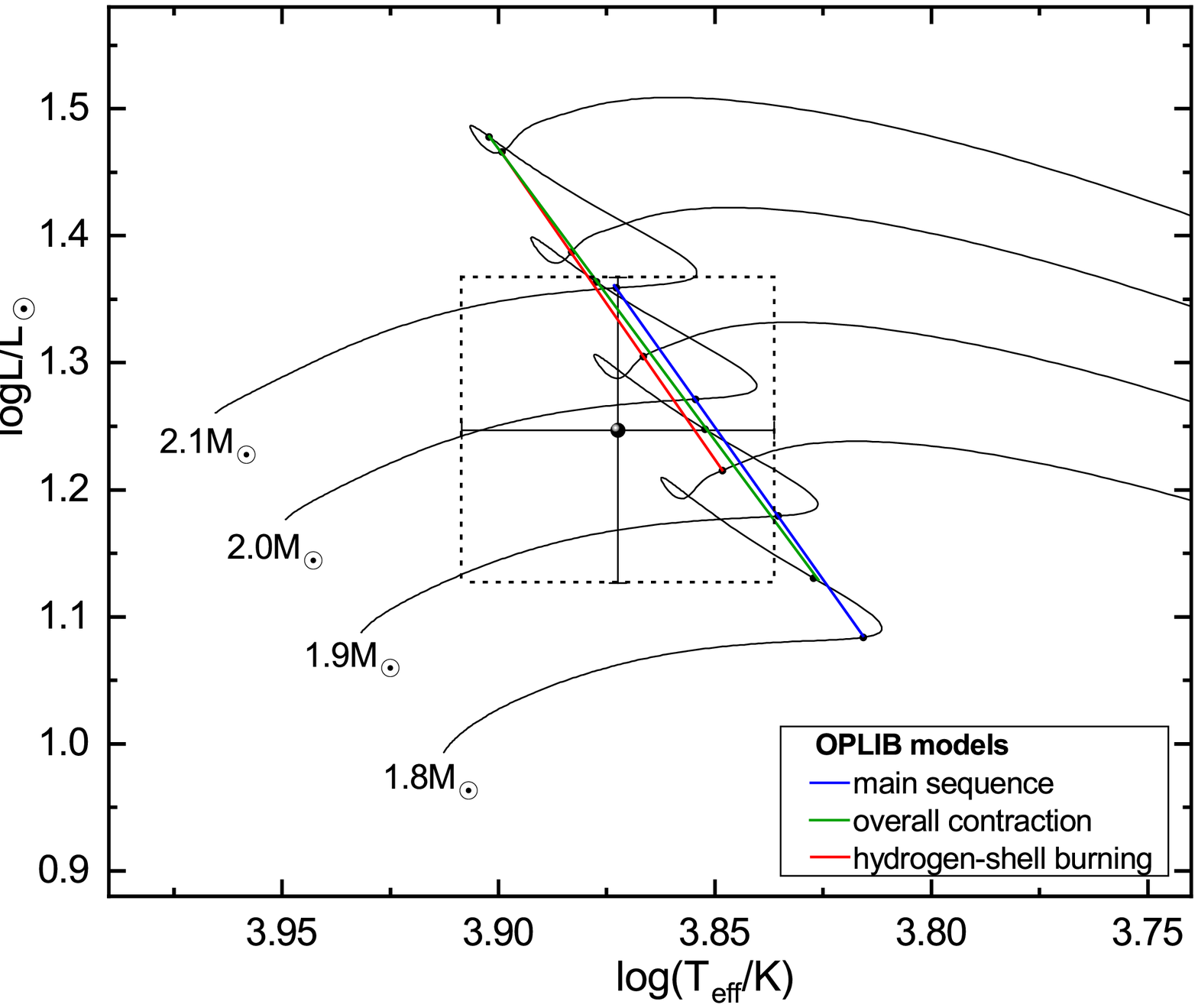}
\caption{The HR diagrams with the observed error box of BP Pegasi and constant frequency lines of the dominant frequency $\nu_1=9.128797$\,d$^{-1}$ as the radial fundamental mode.
Evolutionary models were computed with the OPAL (left), OP (middle) and OPLIB (right) opacity tables.
The other adopted parameters were: the initial hydrogen abundance $X_0=0.70$,
metallicity $Z=0.020$, initial rotational velocity $V_{\rm rot,0}=25~\kms$, the mixing length parameter $\alpha_{\rm MLT}=0.5$. The overshooting from the convective core was not included.
In the case of each opacity data, the models in three stages of evolution are able to reproduce the dominant frequency, i.e., main sequence, overall contraction and hydrogen-shell burning.}
\label{fig6}
\end{figure*}

\begin{figure}
\includegraphics[clip,width=\columnwidth,height=59mm]{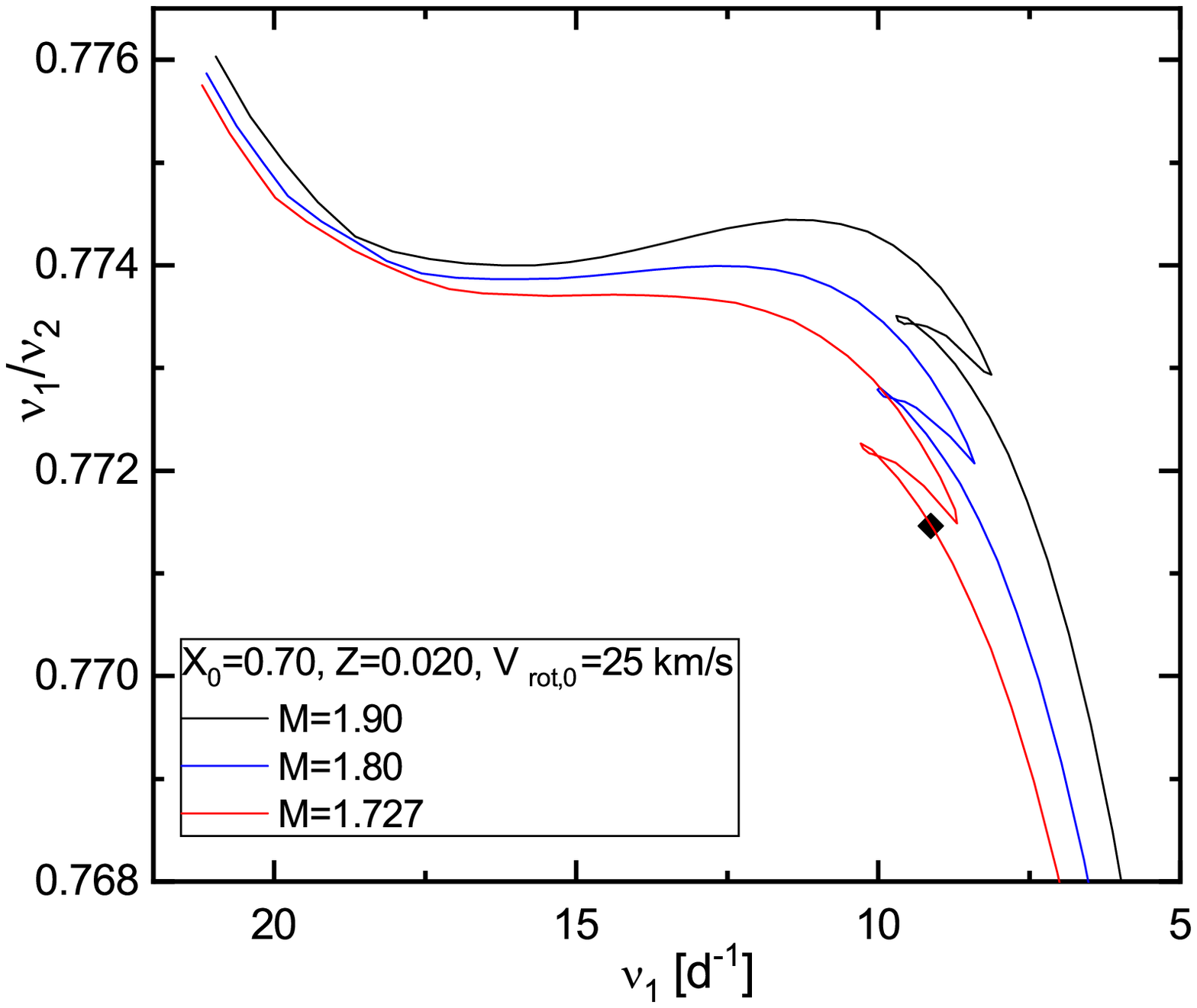}
\includegraphics[clip,width=\columnwidth,height=59mm]{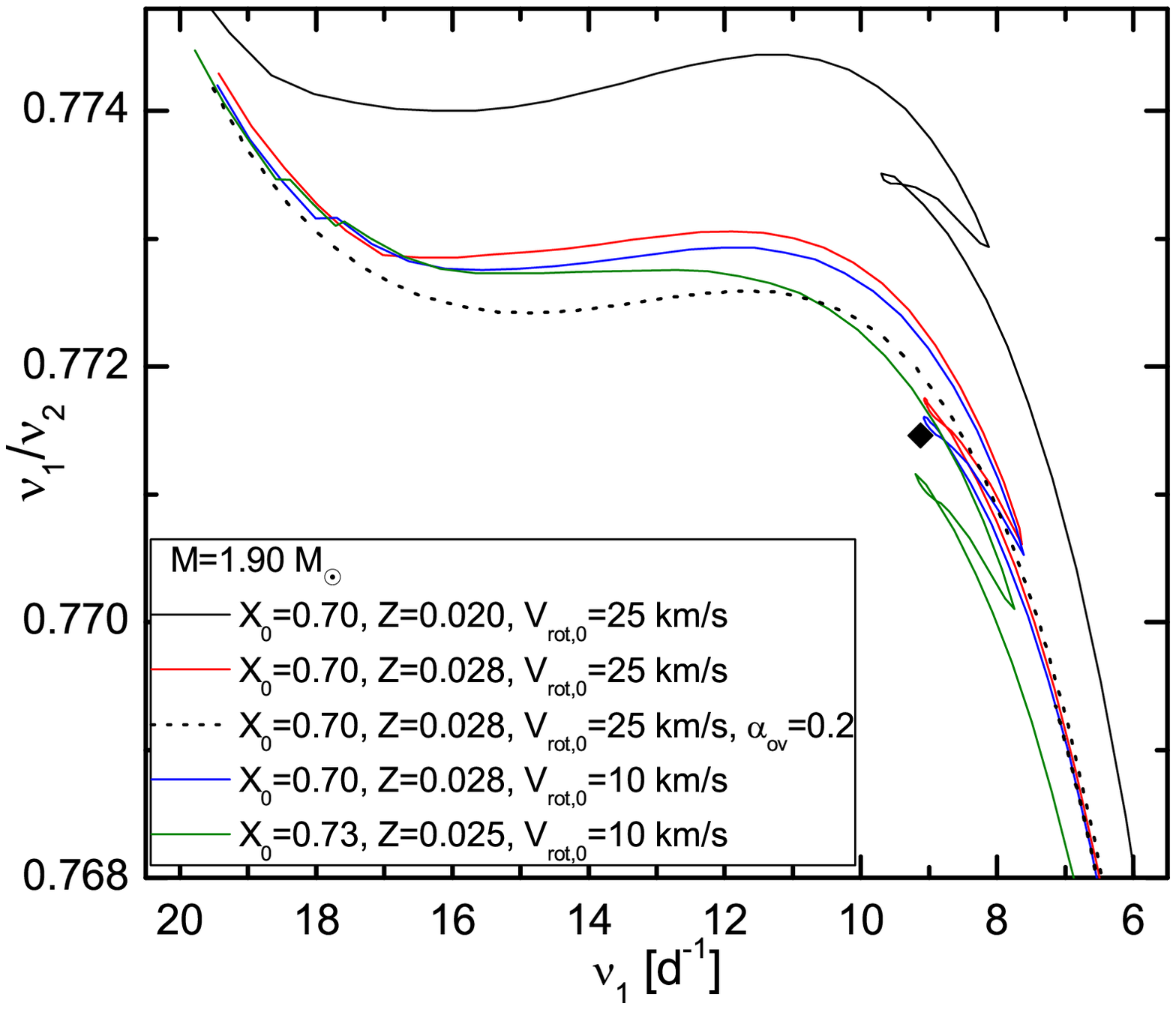}
	\caption{The Petersen diagrams spanning from the zero-age main sequence to the post-main sequence phase of evolution.
The observed value of $(\nu_1,~\nu_1/\nu_2)$ is indicated by the black diamond. There is shown the effect of mass (the top panel) and the effects of hydrogen abundance, metallicity, convective overshooting and rotational velocity (the bottom panel).
The models were computed with the OPAL tables and the mixing length parameter $\alpha_{\rm MLT}=0.5$.}
	\label{fig7}
\end{figure}

Next, we included the second frequency to the seismic modelling. We started with the OPAL data.
The fitting of the two radial modes is best illustrated
on the Petersen diagrams. The examples are given in Fig.\,7. In the top panel we show the effect of mass whereas in the bottom panel
the effects of metallicity, convective overshooting, initial hydrogen abundance and rotation are presented. As one can see the lower the mass the lower the frequency ratio $\nu_1/\nu_2$. From the bottom panel of Fig.\,7, we can draw the following conclusions:
1) the higher the metallicity the lower the value of $\nu_1/\nu_2$, 2) adding overshooting from the convective core decreases the frequency ratio,
3) the higher the hydrogen abundance, the lower the value of $\nu_1/\nu_2$,
and 4) the higher the rotational velocity, the greater the value of $\nu_1/\nu_2$.\\
The effect of rotation on the Petersen diagram has been studied more extensively by 
\citet{Suarez2006}.

The first important result is that with $\alpha_{\rm MLT}=0.5$ only post-main sequence seismic models 
 have the luminosities
that reach the minimum observed value of BP Peg. Secondly, the fitting of the two radial modes gives a strict relation 
between the mass and metallicity. Thirdly, adding convective overshooting moves seismic models away from the error box towards
lower effective temperatures and luminosities. Seismic models with $\alpha_{\rm ov}\ne 0$ are mostly in the main-sequence phase
and have too low luminosities.
In the top panel of Fig.8, we put the OPAL seismic models in the HSB phase on the mass-metallicity diagram.
We included only HSB models without overshooting because they have the highest luminosities reaching the error box.

With the OPAL opacities, we got $Z\in (0.0155, 0.028)$ for the initial hydrogen abundance $X_0=0.70$
and $Z\in (0.013, 0.024)$ for $X_0=0.73$. The lower value of $Z$ is limited by the observed value of the luminosity
whereas the upper limit of $Z$ results directly from the fitting to the two radial modes.
For higher metallicity than $Z_{\rm max}$ for a given $X_0$ there is no model reproducing the observed frequencies.

Then, we repeated seismic modelling with the OP and OPLIB opacity data. It turned out that for $\alpha_{\rm MLT}=0.5$
in the wide range of $X_0,~Z$ there are no models fitting the two frequencies within the error box of BP Peg.
The enormous opacity effect when fitting the two radial modes was also demonstrated by 
\citet{Lenz2007}, who performed seismic modelling for the $\delta$ Sct star 44\,Tau. 
The positions of the seismic models of BP Peg on the HR diagram computed with the OPAL, OP and OPLIB data are shown
in the bottom panel of Fig.\,8. As one can see all OP and OPLIB seismic models are far too cool and too less luminous.
We will examine this rather unexpected result in more detail in the next subsection.
\begin{figure}
\includegraphics[clip,width=\columnwidth,height=118mm]{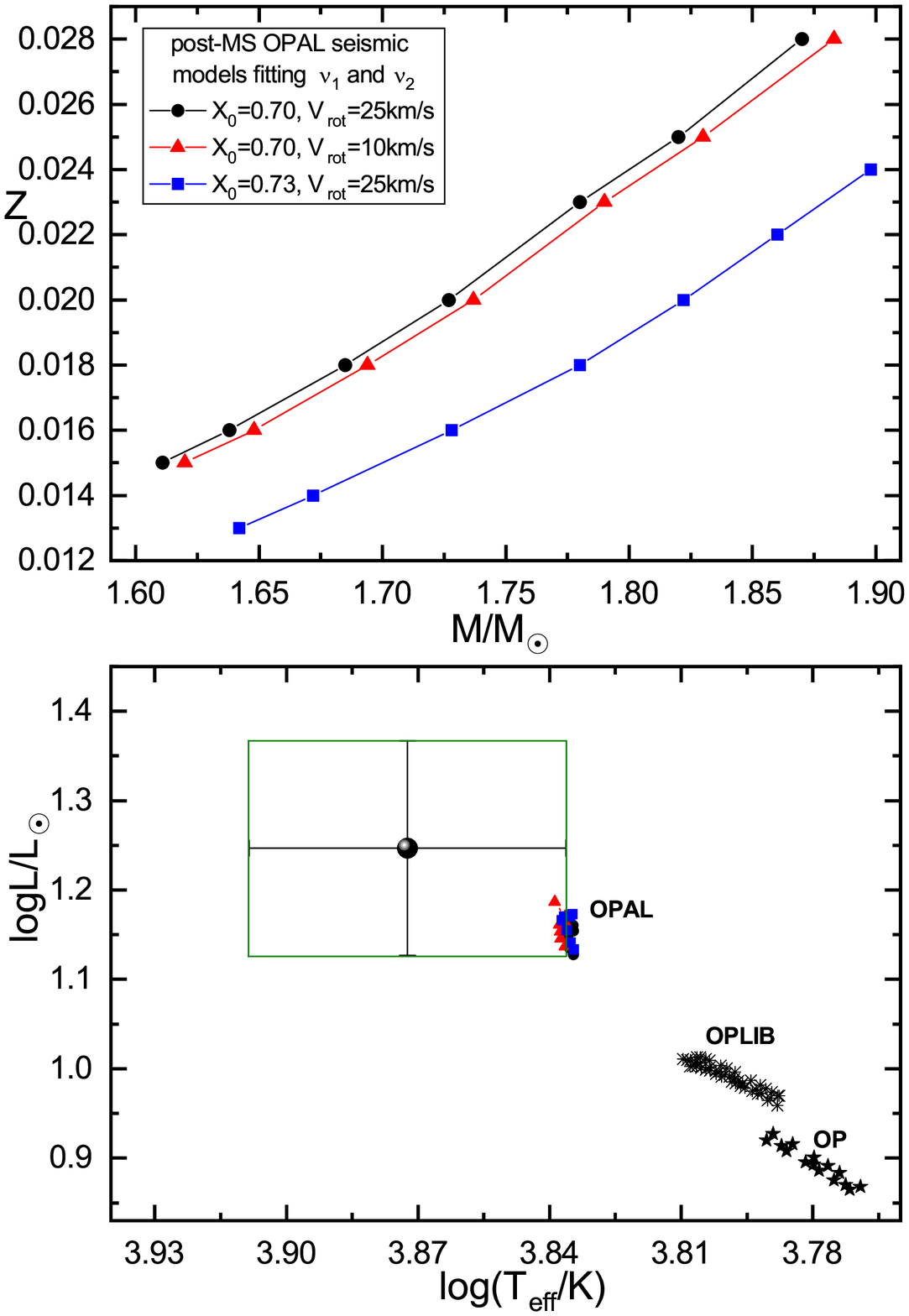}
	\caption{The top panel: the OPAL post-MS seismic models in the HSB phase fitting the two frequencies of BP Peg as the radial fundamental
and first overtone modes on the diagram ''metallicity $Z$ vs. mass $M$''. There is also shown the effect of the initial
hydrogen abundance $X_0$ and initial rotational velocity $V_{\rm rot,0}$. The mixing length parameter was $\alpha_{\rm MLT}=0.5$.
The bottom panel: the position of these models on the HR diagram with the observed error box of BP Peg. There are also shown
positions of the OP and OPLIB seismic models in the HSB phase.}
	\label{fig8}
\end{figure}

\subsection{The effect of $\alpha_{\rm MLT}$}

Till now we made seismic modelling assuming the one value of the mixing length parameter in the envelope, i.e., $\alpha_{\rm MLT}=0.5$.
The natural question arrises: what is the effect of this parameter on models fitting the two radial modes?
The effect of $\alpha_{\rm MLT}$ is illustrated in Fig.\,9. The figure shows the Petersen diagram, with the run of $\nu_1/\nu_2~vs~\nu_1$
for the OPAL model with $M=1.73~{\rm M}_{\sun}$ adopting $\alpha_{\rm MLT}=0.5$ and 1.8, and the two values of metallicity $Z=0.020$ and 0.015.
\begin{figure}
\includegraphics[clip,width=\columnwidth,height=65mm]{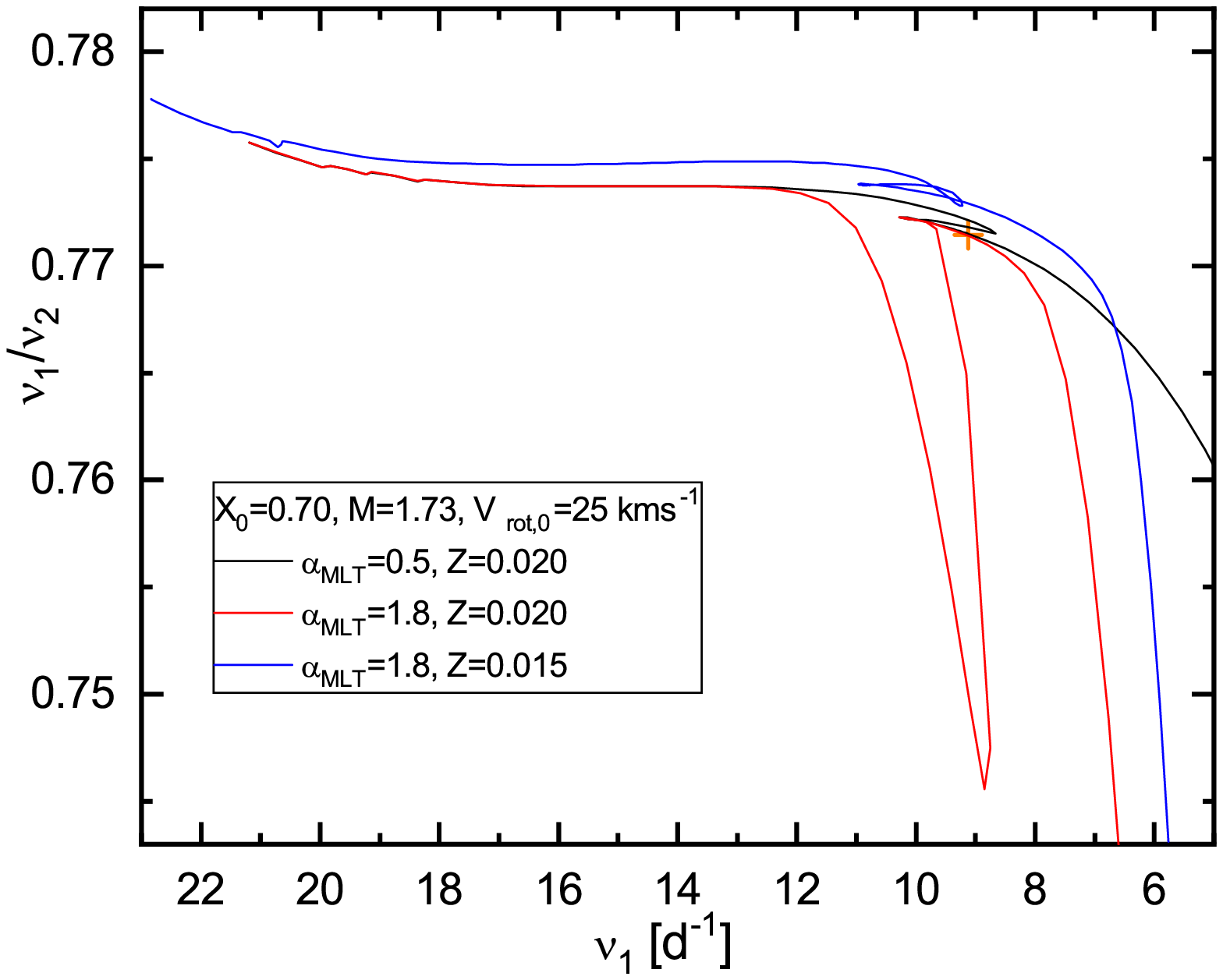}
	\caption{The Petersen diagram for the OPAL model with the mass $M=1.73~{\rm M}_{\sun}$. There is shown the effect
of the mixing length parameter on the frequency ratio $\nu_1/\nu_2$ and the effect of metallicity
for $\alpha_{\rm MLT}=1.8$.}
	\label{fig9}
\end{figure}

In the case of model with $\alpha_{\rm MLT}=1.8$ and $Z=0.02$, we have a deep decrease of $\nu_1/\nu_2$ when
the Terminal Age Main Sequence (TAMS) is approaching. Then, $\nu_1/\nu_2$ increases in the overall contraction phase
reaching the same maximum value as in the case of $\alpha_{\rm MLT}=0.5$. Next, it goes down steeply in hydrogen-shell burning phase. As one can see, this effect depends strongly on the metallicity and for $Z=0.015$
there is no  sharp drop in the frequency ratio.
The sudden drop of $\nu_1/\nu_2$ for $\alpha_{\rm MLT}=1.8$ and $Z=0.020$ is not caused by our assumption about the frozen convection during pulsations
as one could guess. This is because the adiabatic and nonadiabatic  frequencies of the radial fundamental
and first overtone modes are the same up to the forth decimal place.
The reason of such behaviour of $\nu_1/\nu_2$ is the change of the internal structure of cooler models for larger values of $\alpha_{\rm MLT}$,
in particular around the local opacity bumps.  In Fig.\,10, we show the 3D plot of the mean opacity as a function of depth,
expressed by the temperature and density, for $X_0=0.70,~Z=0.020$. The dotted and solid lines correspond to the $M=1.73~{\rm M}_{\sun}$
models with $\alpha_{\rm MLT}=0.5$ and 1.8, respectively. The blue, red and black lines mark the $M=1.73~{\rm M}_{\sun}$ models
near the end of MS (no.\,125), on TAMS (no.\,158) and right after TAMS (no. 170), respectively. 
\begin{figure*}
	\includegraphics[width=175mm,clip]{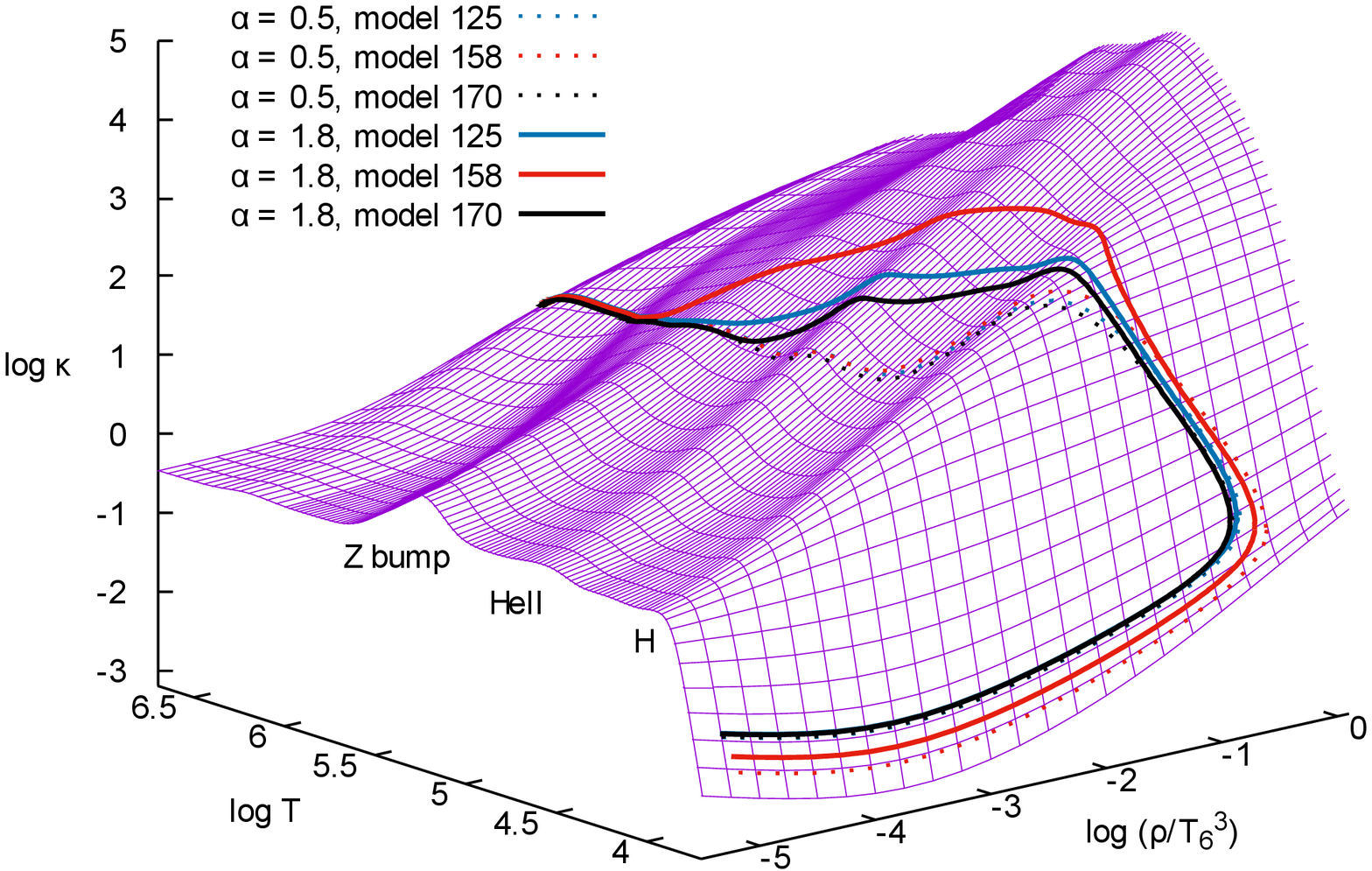}
\caption{The mean Rosseland opacities $\kappa$ for $X_0=0.70,~Z=0.020$ and the OPAL data.
 Dotted and solid lines depict the models with a mass $M=1.73~{\rm M}_{\sun}$ computed with the mixing length
 parameter $\alpha_{\rm MLT}=0.5$ and 1.8, respectively. The blue, red and black lines mark the models near the end of MS,
 on TAMS and right after TAMS, respectively.}
	\label{fig10}
\end{figure*}

For a wide range of $X_0,~Z$, it appeared that increasing the value of $\alpha_{\rm MLT}$ in the OP and OPLIB models,
brings them closer to the error box. For $\alpha_{\rm MLT}\ge 2.0$ it is possible to get seismic models with the accepted effective
temperatures and luminosities.

In Table\,5, we present examples of seismic models in the three phases of evolution computed with the OPAL, OP and OPLIB opacity tables, adopting  the two values of the mixing parameter $\alpha_{\rm MLT}=0.5$ and 2.0.
For $\alpha_{\rm MLT}=0.5$ only the OPAL seismic model in the HSB phase has the effective temperature and  luminosity consistent with the observed values. Moreover, in the case of the OP and OPLIB tables there are no models in the MS and OC phases reproducing the two frequencies of BP Peg at all. 
For $\alpha_{\rm MLT}=2.0$, all phases of evolution for each opacity data are possible.  Although, the values of $T_{\rm eff}$ and $\log L/L_{\sun}$ are often marginally consistent with observational determinations.  

\begin{table*}
\centering
\caption{The parameters of the seismic models that reproduce $\nu_1=9.128797$\,d$^{-1}$ as the radial fundamental mode
and $\nu_2=11.83315$\,d$^{-1}$ as the radial first overtone, for the three opacity data and two values of the
mixing length parameter $\alpha_{\rm MLT}=0.5,~2.0$. The other parameters are: $X_0=0.70,~Z=0.020,~\alpha_{\rm ov}=0.0$. }
\begin{tabular}{ccccccccccccc}
\hline
\multicolumn{4}{c}{OPAL} \\
\hline
$\alpha_{\rm MLT}$ & phase & $M/{\rm M}_{\sun}$ & $\log T_{\rm eff}$ & $\log L/{\rm L}_{\sun}$ & $R/{\rm R}_{\sun}$ & age [Gyr] & $\nu(p_2)$ [d$^{-1}$] & $\nu(p_1)/\nu(p_2)$  & $\eta(p_1)$ & $\eta(p_2)$ \\
\hline
               0.5 &   MS  &      1.674   &                 3.7918     &  0.964            &    2.65      &    1.57   & 11.833146  &       0.77146       &    0.073    & 0.022    \\
               0.5 &   OC  &      1.696   &                 3.8064     &  1.027            &    2.66      &    1.58   & 11.833157  &       0.77146       &    0.092    & 0.054    \\
              0.5 &  HSB &      1.726   &                 3.8362     &  1.152            &    2.68      &    1.52   & 11.833146  &       0.77146       &    0.101    & 0.091    \\
\hline
               2.0 &   MS  &      1.860   &                 3.8342     &  1.149            &    2.69      &    1.10   & 11.833217  &       0.77146       &    0.439    & 0.459    \\
               2.0 &   OC  &      1.797   &                 3.8365     &  1.152            &    2.67      &    1.35   & 11.833283  &       0.77145       &    0.338    & 0.337    \\
               2.0 & HSB  &      1.745   &                 3.8414     &  1.169            &    2.67      &    1.48   & 11.833351  &       0.77145       &    0.198    &  0.195 \\
\hline
\multicolumn{4}{c}{OP} \\
\hline
$\alpha_{\rm MLT}$ & phase & $M/{\rm M}_{\sun}$ & $\log T_{\rm eff}$ & $\log L/{\rm L}_{\sun}$ & $R/{\rm R}_{\sun}$ & age [Gyr] & $\nu(p_2)$ & $\nu(p_1)/\nu(p_2)$  & $\eta(p_1)$ & $\eta(p_2)$ \\
\hline
            0.5 & HSB  & 1.501    &      3.7994          &     0.920    &  2.55    &  2.36  & 11.833146  &   0.77146      &   0.076     &   0.023   \\
\hline
               2.0 &   MS  &  1.834  &       3.8302          &    1.123     &  2.66    &  1.19    &  11.831851  &   0.77154  &  0.443  &  0.432  \\
               2.0 &   OC  &  1.777  &       3.8311          &    1.119     &  2.64    &  1.41    &  11.831579  &   0.77156  &  0.458  &  0.437  \\
            2.0 & HSB  &  1.685  &       3.8322          &    1.111     &  2.60    &  1.66    &  11.833919  &   0.77141  &  0.480  &  0.446  \\
\hline
\multicolumn{4}{c}{OPLIB} \\
\hline
$\alpha_{\rm MLT}$ & phase & $M/{\rm M}_{\sun}$ & $\log T_{\rm eff}$ & $\log L/{\rm L}_{\sun}$ & $R/{\rm R}_{\sun}$ & age [Gyr] & $\nu(p_2)$ & $\nu(p_1)/\nu(p_2)$  & $\eta(p_1)$ & $\eta(p_2)$ \\
\hline
            0.5 & HSB  &  1.577   &     3.8044           &    0.999     & 2.60     &  1.94  &  11.833146 &  0.77146  &  0.094    &  0.066  \\
\hline
               2.0 &   MS  &  1.856  &      3.8324          &   1.138     &  2.68    & 1.10 &  11.833146 &  0.77146      &  0.416    &  0.439   \\
               2.0 &   OC  &  1.805  &      3.8332          &   1.135     &  2.66    & 1.28 &  11.833146 &  0.77146      &  0.430    &  0.455   \\
            2.0 & HSB  &  1.708  &      3.8350          &   1.130     &  2.63    & 1.52 &  11.833140 &  0.77146      &  0.414    &  0.427   \\
\hline
\end{tabular}
\end{table*}


\subsection{Constraints on convection from the parameter $f$}
The potential of the parameter $f$ to obtain the information on the efficiency of convective transport has been demonstrated in many works \citep{JDD2003,JDD2020, JDD2005, JDD2021}.
Let us recall that this parameter describes the relative amplitude of the radiative flux  variations at the photosphere level for a given pulsational mode (see Sect.\,4).
The theoretical value of $f$ can be derived only in the framework of nonadiabatic theory of stellar pulsations as it is complex because, besides the amplitude
of the flux variations, we need also the phase shift between the flux and radius variations.
The parameter $f$ is very sensitive to the subphotospheric condition and, in the case of $\delta$ Scuti pulsators, the envelope convection greatly modifies
the value of $f$. This fact has profound consequences for the mode identification because the parameter $f$ enters the expression for the photometric amplitudes.

On the other hand, the empirical value of $f$ for a given mode can be obtained from the observed photometric amplitudes and phases simultaneously
with the identification of the mode degree $\ell$. In that way, the determination of $\ell$ is independent of the  treatment of convection whereas
the empirical values of $f$ can be  directly compared with the theoretical predictions for different values of the mixing length parameter $\alpha_{\rm MLT}$,
that is for different efficiency of convective transport in the envelope.
\begin{figure}
	\includegraphics[clip,width=\columnwidth]{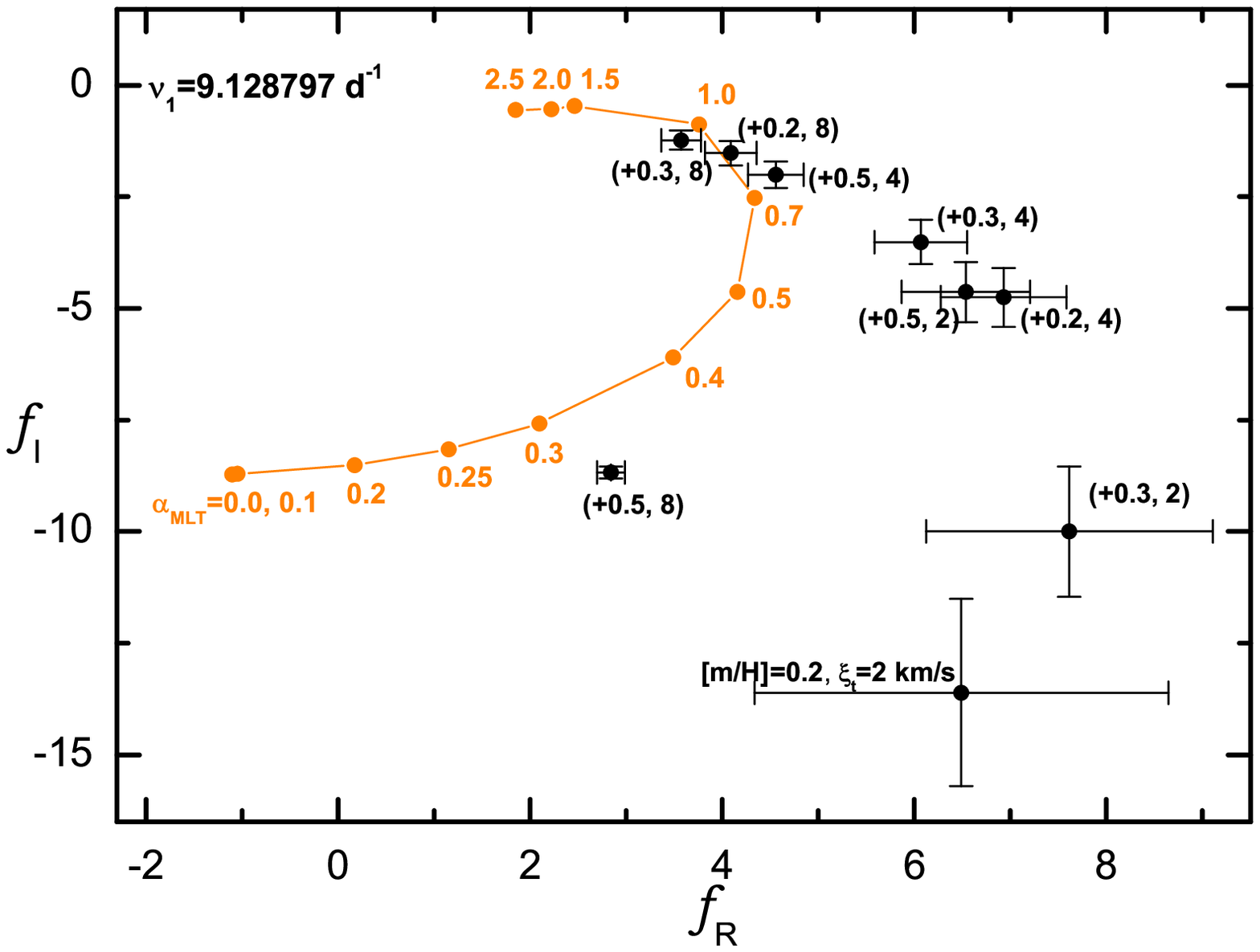}
	\caption{A comparison of the theoretical and empirical values of $f$ for the fundamental radial mode of BP Pegasi.
		The theoretical values of $f$ are for the OPAL seismic model with the parameters: $X_0=0.70,~Z=0.023,~M=1.79~{\rm M}_{\sun}$, ~$\log T_{\rm eff}\approx 3.835$,~$\log L/L_{\sun}\approx 1.164$ and
		twelve value of $\alpha_{\rm MLT}$ from 0 to 2.5. The empirical values of $f$ were determined with the NEMO model atmospheres
		for different atmospheric metallicities [m/H] and microturbulent velocities $\xi_t$.}
	\label{fig11}
\end{figure}

Such a comparison for the radial fundamental mode of BP Pegasi is presented in Fig.\,11 on the complex plane. The theoretical values of $f$ are for the OPAL
seismic model reproducing the two radial-mode frequencies and twelve values of the mixing length parameter $\alpha_{\rm MLT}$.
The model parameters are: $X_0=0.70,~Z=0.023,~M=1.79~{\rm M}_{\sun}$, ~$\log T_{\rm eff}\approx 3.8365$,~$\log L/L_{\sun}\approx 1.164$.
The effective temperature and luminosity are approximate as they may slightly differ between 
different values of $\alpha_{\rm MLT}$.
The empirical values of $f$  for the second frequency of BP Peg have too large errors (of the order of 3 to 5) to be useful for getting
reliable constraints.

As one can see from Fig.\,11, the agreement between the theoretical and empirical values of $f$ for $\nu_1$ can be achieved for the mixing parameters in the range of about $\alpha_{\rm MLT}\in (0.7,~1.0)$
if the atmospheric metallicity and microturbulent velocity are ([m/H],\,$\xi_t$) =(0.2,\,8$~\kms$), ([m/H],\,$\xi_t$) =(0.3,\,8$~\kms$) or ([m/H],\,$\xi_t$) =(0.5,\,4$~\kms$).
As we have checked,  the lower values of [m/H] and $\xi_t$ are definitely  excluded. Moreover,  the lower the metallicity and microturbulent velocity the larger the errors of $f$.  This means  that lower values of  [m/H] and $\xi_t$ give a much worse fit between the calculated and observed photometric amplitudes and phases.

As one can see, the obtained range of $\alpha_{\rm MLT}$ is quite narrow
and amounts to about (0.7,\,1.0). Is this conclusion valid for other seismic models? To answer this question as fully as possible,
we will perform more extensive seismic analysis in the next Section.

\section{Asteroseismic modelling with Monte Carlo-based Bayesian analysis}
In the previous section, we showed the results of seismic modelling of BP Peg considering only some sets of parameters
for the three sources of the opacity data, OPAL, OP and OPLIB. These results allowed to draw the two important conclusions.
Firstly, main-sequence seismic models, reproducing the two radial-mode frequencies, are much less likely because
their luminosities  are too low compared to the observed values. Secondly, only with the OPAL tables it is possible
to construct  seismic models with the mixing length parameter $\alpha_{\rm MLT}<2$ and the parameters
($T_{\rm eff},~L/L_{\sun}$) consistent with the observed determinations.

In this section, we present the more advanced approach using the Bayesian analysis based on the Monte Carlo method.
For the reasons mentioned above, we will consider only the evolutionary phases of overall contraction and hydrogen-shell burning,
and we will use the OPAL opacities.

The analysis is based on the Gaussian likelihood function defined as \citep[e.g.,][]{2005A&A...436..127J, 2006A&A...458..609D, 2017MNRAS.467.1433R,Jiang2021}
%
$${\cal L}(E|{\mathbf H})=\prod_{i=1}^n \frac1{\sqrt{2\pi\sigma_i^2}} \cdot
{\rm exp} \left( - \frac{  ({\cal O}_i-{\cal M}_i)^2}{2\sigma_i^2}  \right), \eqno(3)$$
where ${\mathbf H}$ is the hypothesis that represents adjustable model and theory parameters that in our case are:
mass $M$, initial hydrogen abundance $X_0$, metallicity $Z$, rotation $V_{\rm rot}$,
convective overshooting parameter $\alpha_{\rm ov}$ and the mixing length parameter $\alpha_{\rm MLT}$.
The evidence $E$ represents the calculated observables ${\cal M}_i$, e.g.,  the effective temperature $T_{\rm eff}$, luminosity $L/L_{\sun}$, pulsational frequencies,  that can be directly compared with the observed parameters ${\cal O}_i$ determined with the errors $\sigma_i$.

Here we used the following observations: effective temperature $T_{\rm eff}$, luminosity $L/L_{\sun}$,
the frequencies of the two radial modes $\nu_1$ and $\nu_2$, and the parameter $f$ of the dominant mode.
Then, we made a huge number (about 80\,000) of simulations to maximize the likelihood function given in Eq.\,(3)
in order to constrain the model parameters.
.
\subsection{Simulations without the parameter $f$}
In the first grid of simulation, we fixed the mixing parameter $\alpha_{\rm MLT}=0.5$. The other five parameters of models, i.e., mass, initial rotational velocity, metallicity, hydrogen abundance and overshooting parameter were randomly generated during simulation. For each randomly selected set of parameters we calculated evolutionary and pulsational models.
Then, we chose the models that had the frequency of the radial fundamental mode fitting  the observed frequency $\nu_1$.
Thus, the models, we considered, reproduced exactly the dominant frequency.  The OC and HSB models were considered separately, that is we ran independent simulations for each of them.

For $X_0$ we assumed a beta function $B(2,2)$ as a prior probability, since we wanted to restrict its value to the reasonable 
range, i.e., from 0.6 to 0.8 with $X_0=0.7$ as the most probable.  For other parameters we used uninformative priors, i.e., a uniform distribution. 
The values  of the effective temperature and luminosity with uncertainties were given in Sect.\,2. The value of the second frequency 
corresponding to the first overtone mode was taken from Table\,2. However, rather than using the formal error from
 the least square fitting procedure, we  took the Rayleigh resolution
of the ASAS data, as a more realistic measure of the error, that is $\sigma_{\nu}=0.0004$ d$^{-1}$.
\begin{figure*}
	\begin{multicols}{2}
		\includegraphics[width=88mm,clip]{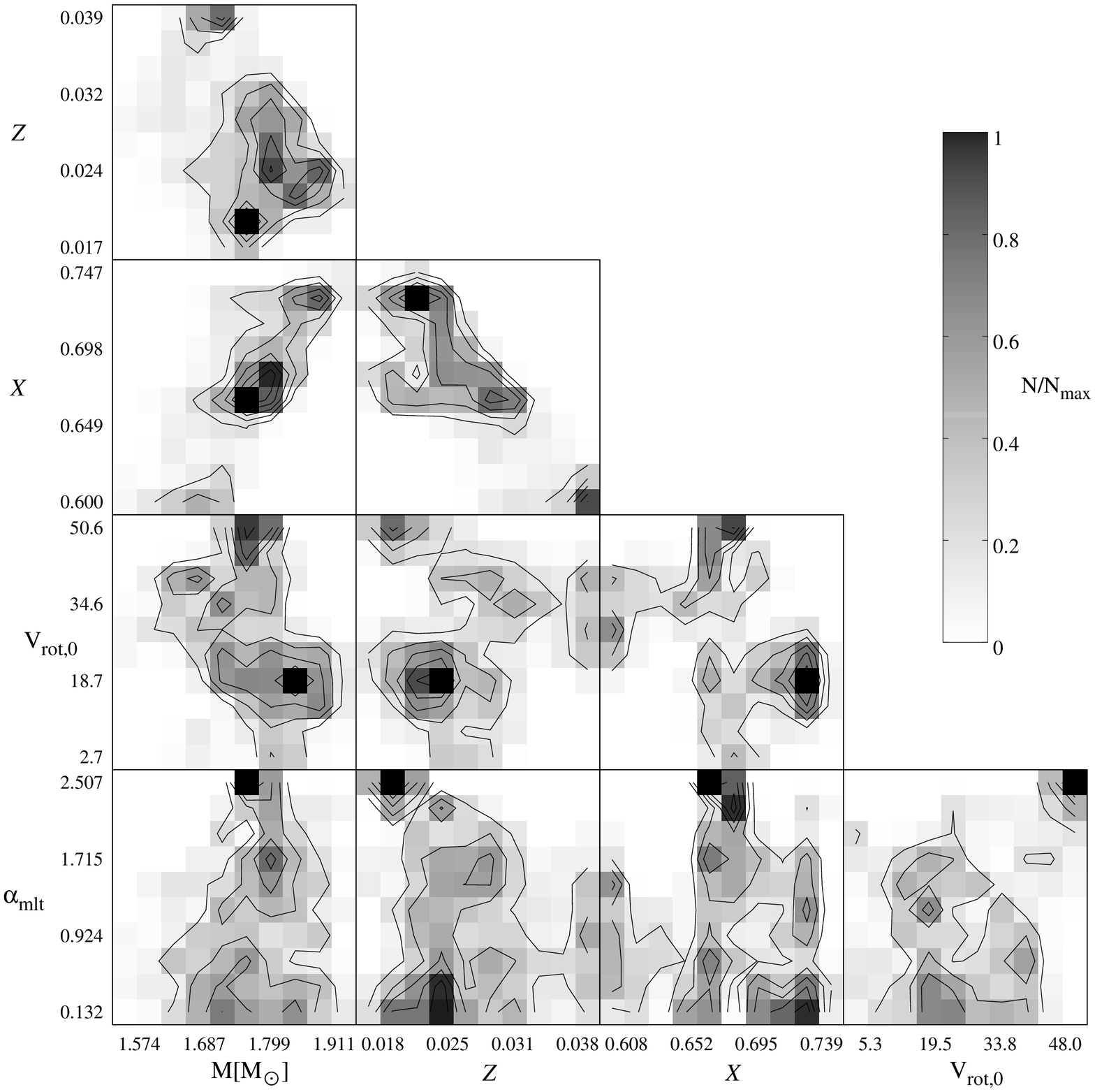}\par
		\includegraphics[width=88mm,clip]{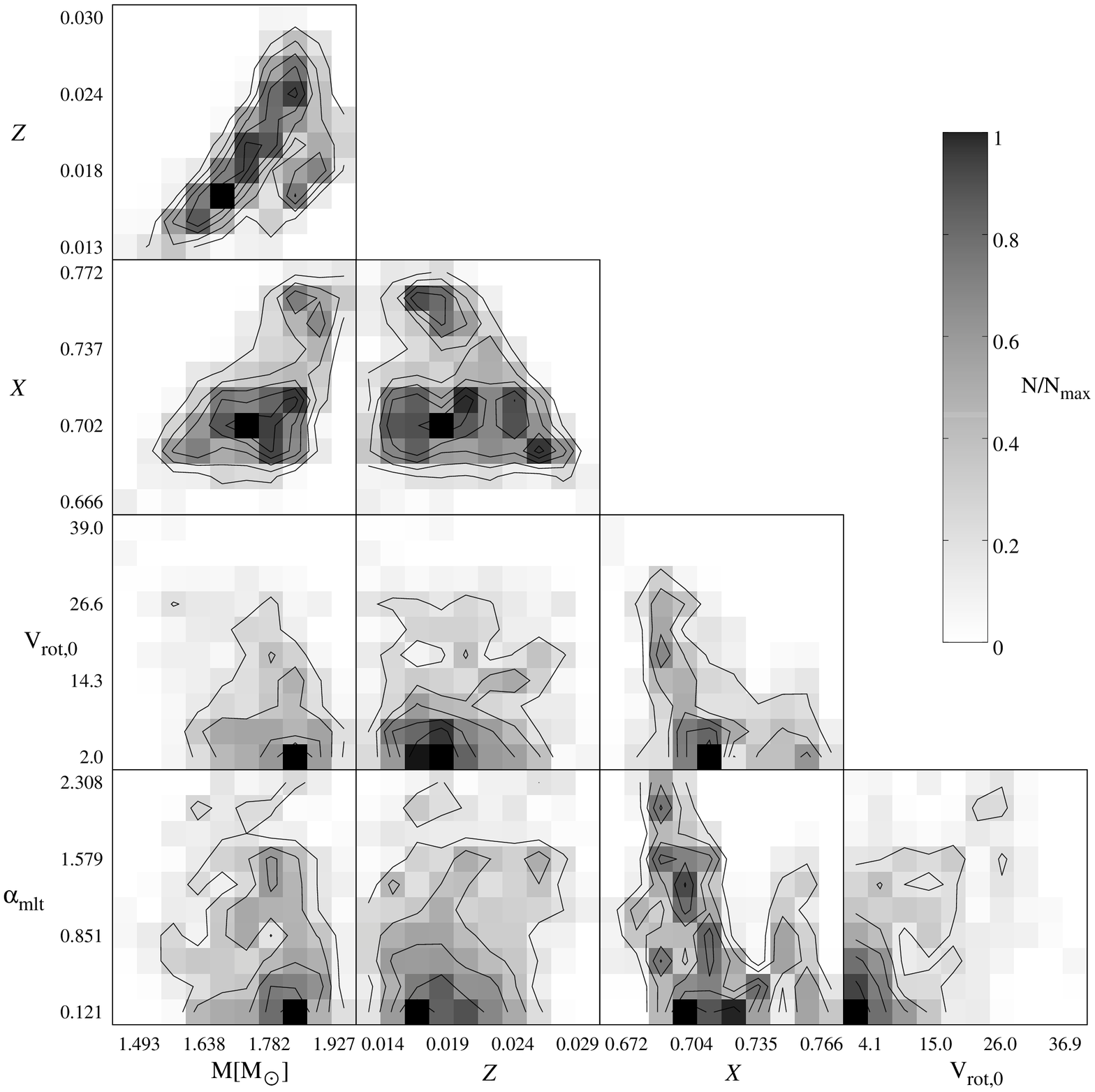}\par
	\end{multicols}
	\begin{multicols}{2}
		\includegraphics[width=88mm,clip]{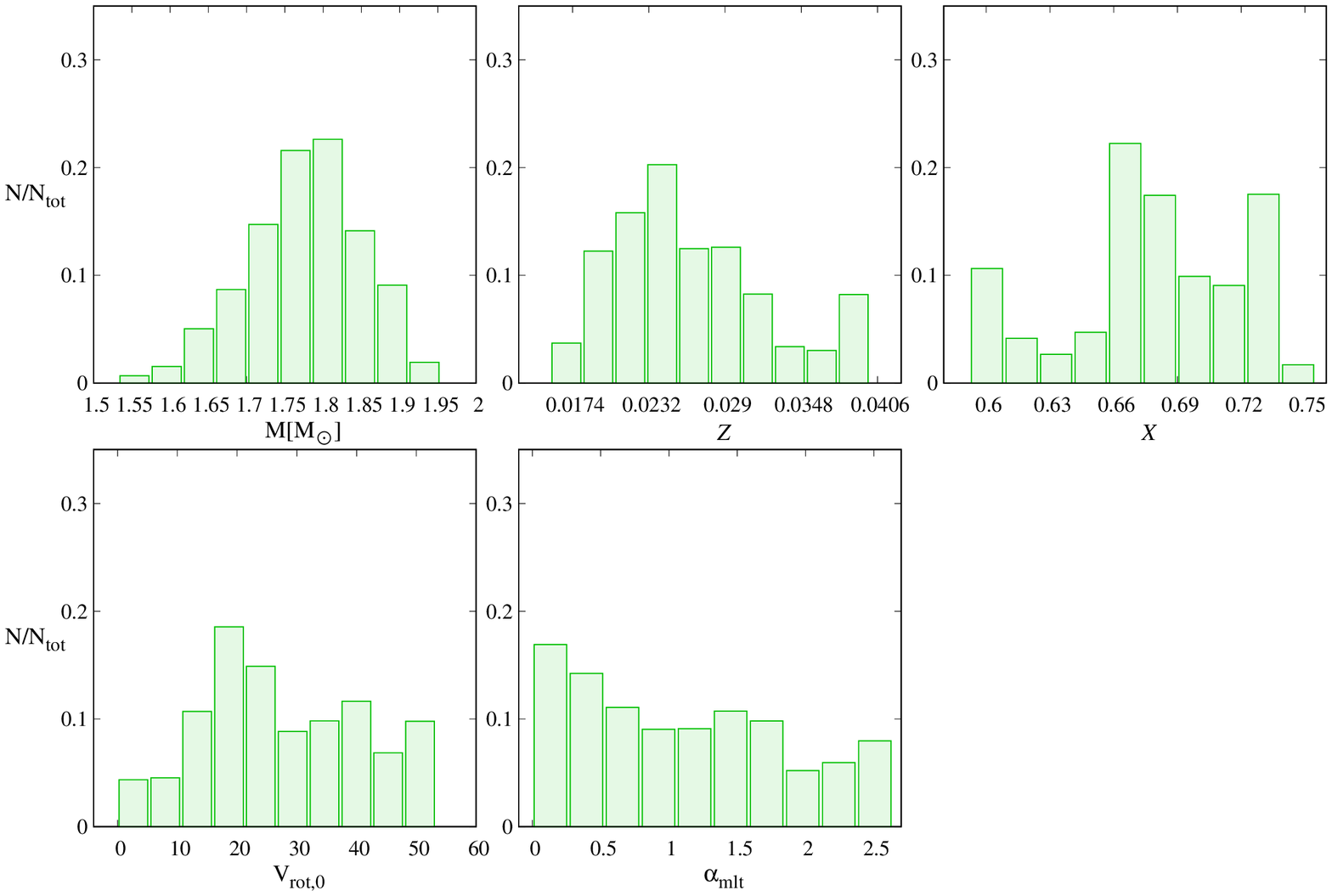}\par
		\includegraphics[width=88mm,clip]{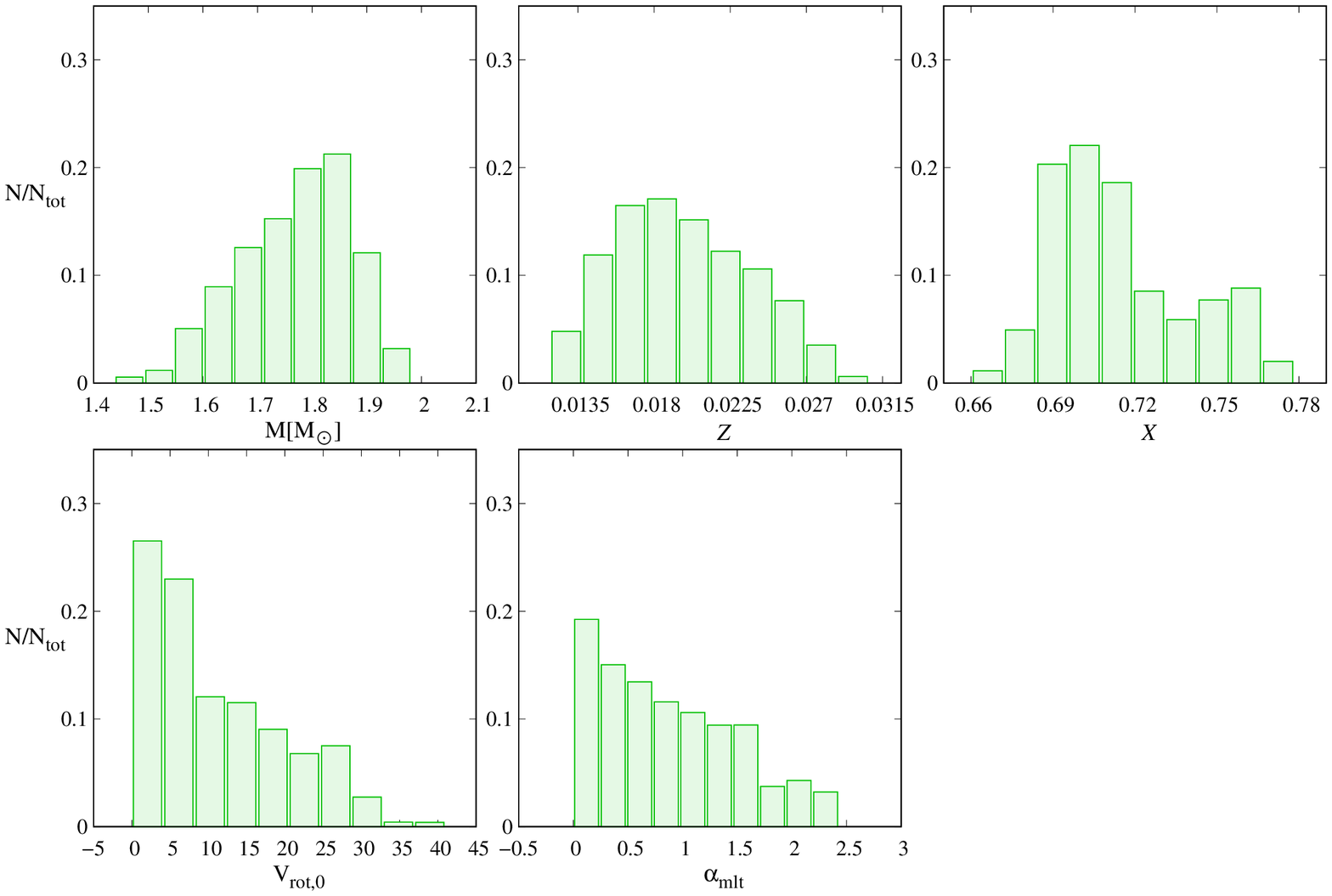}\par
	\end{multicols}
	\caption{Upper panels: the corner plots for the parameters: $M$, $X_0$, $Z$, $V_{\rm rot}$ and $\alpha_{\rm MLT}$.
		Bottom panels: the normalized histograms for the same parameters. The left column corresponds to the OC models, while the right column to the HSB models.}
	\label{fig:mcs:cp1}
\end{figure*}

The important result of our first simulations was that convective overshooting is rather ineffective because  in all runs the parameter $\alpha_{\rm ov}$ tended to zero.  
We already received such a hint in Sect.\,5 where we found that seismic models with $\alpha_{\rm ov}=0.2$ have too low luminosities.
Therefore, in the following runs we fixed $\alpha_{\rm {ov}}=0.0$ while $\alpha_{\rm MLT}$ was henceforth treated as a free parameter.  Thus, we searched for the best values of the parameters:  $M$, $X_0$, $Z$, $V_{\rm rot}$ and  $\alpha_{\rm MLT}$.  
\begin{figure}
	\includegraphics[angle=0, width=0.95\columnwidth]{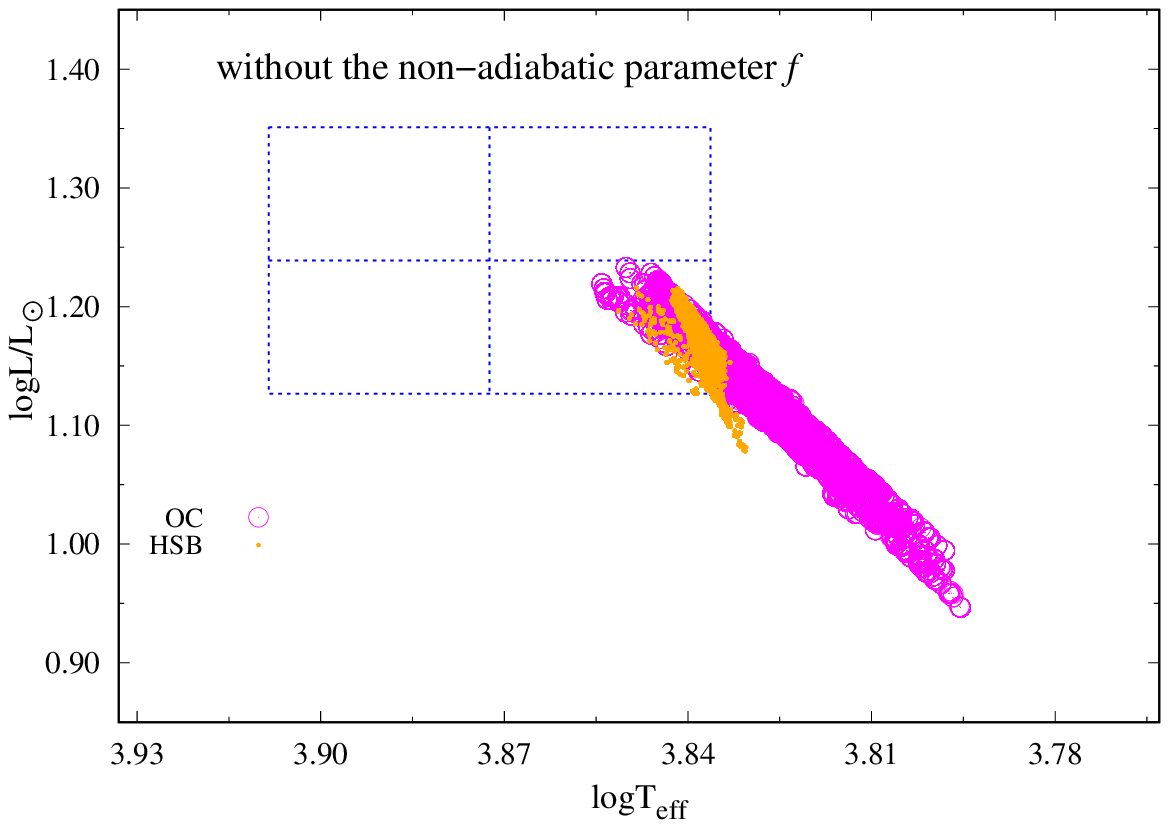}
	\caption{The HR diagram with the position of the seismic models obtained from the simulations.
		The OC  models are marked with magenta circles and the HSB models  with orange dots.}
	\label{fig:mcs:HR1}
\end{figure}

The results of our simulations are shown in Fig.\,\ref{fig:mcs:cp1}. In the upper panels we plotted the corner plots
for $M$, $X_0$, $Z$, $V_{\rm rot}$ and $\alpha_{\rm MLT}$. In the bottom panels we showed the corresponding  histograms. The left column indicates the OC seismic models, while the right column - the HSB seismic models.
The histograms were normalised to 1.0  by the number of all models, thus the numbers on the Y-axis times 100
are the percentage  of models with a given parameter range.

As one can see, for both the OC and HSB models, the mass and metallicity are the best constrained parameters.
The initial  hydrogen abundance $X_0$ for the OC models seems to have two extremes, near $X\approx0.68$ and $X\approx0.73$. The HSB models concentrate around $X\approx0.71$.
The least constrained parameters, in both cases, are $V_{\rm rot}$ and $\alpha_{\rm MLT}$. The models seem to be weakly dependent on them.
The expected values of the parameters from the distributions represented by the histograms in Fig.\,\ref{fig:mcs:cp1}
are given in the first two lines of Table\,6.  The uncertainties were calculated as the square roots of the variance.

All  models from the simulations are also marked in the HR diagram shown in Fig.\,\ref{fig:mcs:HR1}. All of them
concentrate near the bottom right corner of the error box. All HSB models lay within $2\sigma$ of the effective temperature and luminosity whereas most of the OC models are within $3\sigma$ errors.

		\setlength{\tabcolsep}{4pt}
		\begin{table*}
			\begin{center}
				\label{tab}
				\caption{The expected values of the parameters of BP Peg obtained  from the Monte-Carlo simulations. 
					The uncertainties were calculated as the square roots of the variance. The first two lines correspond to the simulations without the parameter $f$ of the radial fundamental mode and the remaining six lines to the simulations with this parameter $f$.
				In the second case, the first column contains the microturbulent velocity in the atmosphere $\xi_t$ whereas the atmospheric metallicity [m/H] were changed consistently with $Z$ given in the fourth column.}
				\begin{tabular}{|c|c|c|c|c|c|c|c|c|c|c|c|c|c|c|}
					\hline
					$\xi_t$ & evolution  & $M/{\rm M}_{\sun}$  & $Z$  & $X_0$  & $V_{\mathrm{rot,0}}$  & $\alpha_{\rm MLT }$  & $\log{T_{\mathrm{eff}}}$  & $\log{L/{\rm L}_{\sun}}$
					& $R/{\rm R}_{\sun}$  & $V_{\mathrm{rot}}$  & $\log{(t/{\rm yr})}$  \\
					 $[\kms]$ & phase &  	 &  & & $[\kms]$ &   &   &  &  &  $[\kms]$ &  \\ \hline
										
					--- &  OC & $1.782(72)$   &  $0.0256(49)$   &  $0.686(35)$   &  $26.5(11.9)$   &  $1.13(68)$   &  $3.8300(120)$   &  $1.132(57)$   &  $2.69(4)$   &  $24.3(10.6)$   &  $9.157(35)$  \\
										
					--- &  HSB &$1.775(94)$   &  $0.0200(39)$   &  $0.715(24)$   &  $11.6(7.8)$   &  $0.95(58)$   &  $3.8386(26)$   &  $1.169(20)$   &  $2.70(5)$   &  $11.0(7.3)$   &  $9.181(16)$  \\ \hline

					2 &  OC & $1.718(40)$   &  $0.0314(50)$   &  $0.643(21)$   &  $36.(17.9)$   &  $0.71(36)$   &  $3.8217(77)$   &  $1.086(35)$   &  $2.66(3)$   &  $33.3(17.4)$   &  $9.169(20)$  \\
										
					2 &   HSB & $1.951(47)$   &  $0.0172(14)$   &  $0.783(9)$   &  $12.4(7.1)$   &  $0.16(10)$   &  $3.8423(13)$   &  $1.214(10)$   &  $2.80(2)$   &  $11.4(6.5)$   &  $9.189(7)$   \\

					4 &   OC & $1.803(23)$   &  $0.0259(24)$   &  $0.698(12)$   &  $36.6(13.2)$   &  $0.58(8)$   &  $3.8276(50)$   &  $1.128(23)$   &  $2.71(1)$   &  $33.8(12.9)$   &  $9.187(13)$  \\
										
					4 &   HSB & $1.805(45)$   &  $0.0286(29)$   &  $0.674(21)$   &  $12.9(9.0)$   &  $0.63(7)$   &  $3.8359(14)$   &  $1.162(10)$   &  $2.71(2)$   &  $12.1(8.4)$   &  $9.157(12)$   \\

					8 &   OC &  $1.921(28)$   &  $0.0211(18)$   &  $0.752(15)$   &  $20.5(10.8)$   &  $0.97(16)$   &  $3.8406(53)$   &  $1.200(24)$   &  $2.78(2)$   &  $18.9(9.8)$   &  $9.183(11)$   \\
					
					8 &   HSB &  $1.618(38)$   &  $0.0179(14)$   &  $0.677(10)$   &  $16.1(7.1)$   &  $0.71(8)$   &  $3.8367(9)$   &  $1.136(8)$   &  $2.62(2)$   &  $15.4(6.7)$   &  $9.185(8)$   \\
					
					\hline
				\end{tabular}
			\end{center}
		\end{table*}
		

\subsection{Simulations with the parameter $f$}
In the next step we added more constrains to our simulations. We used the non-adiabatic $f$-parameter of the radial fundamental mode.  This parameter is complex so we have two more constraints. As has been earlier mentioned, the theoretical value of $f$ in $\delta$ Sct stellar models strongly depends on the convection efficiency in the envelope. 
Thus, from a comparison of the theoretical and empirical values of $f$ it is feasible to obtain information on the mixing length parameter $\alpha_{\rm MLT}$.

We calculated the second grid of models for which the likelihood function was modified by multiplying Eq.\,(3)  by two additional factors that contain the real ($f_R$) and imaginary part ($f_I$) of the parameter $f$ of the radial fundamental mode.
Now the results depend on the model atmosphere. As in Sect.\,5,  we used the NEMO models. We made simulations  for a discrete values of the microturbulent velocity $\xi_t=2,~4,~8~\kms$ and the atmospheric metallicity [m/H] was changed consistently with the current value of $Z$.

\begin{figure*}
	\begin{multicols}{2}
		\includegraphics[width=88mm,clip]{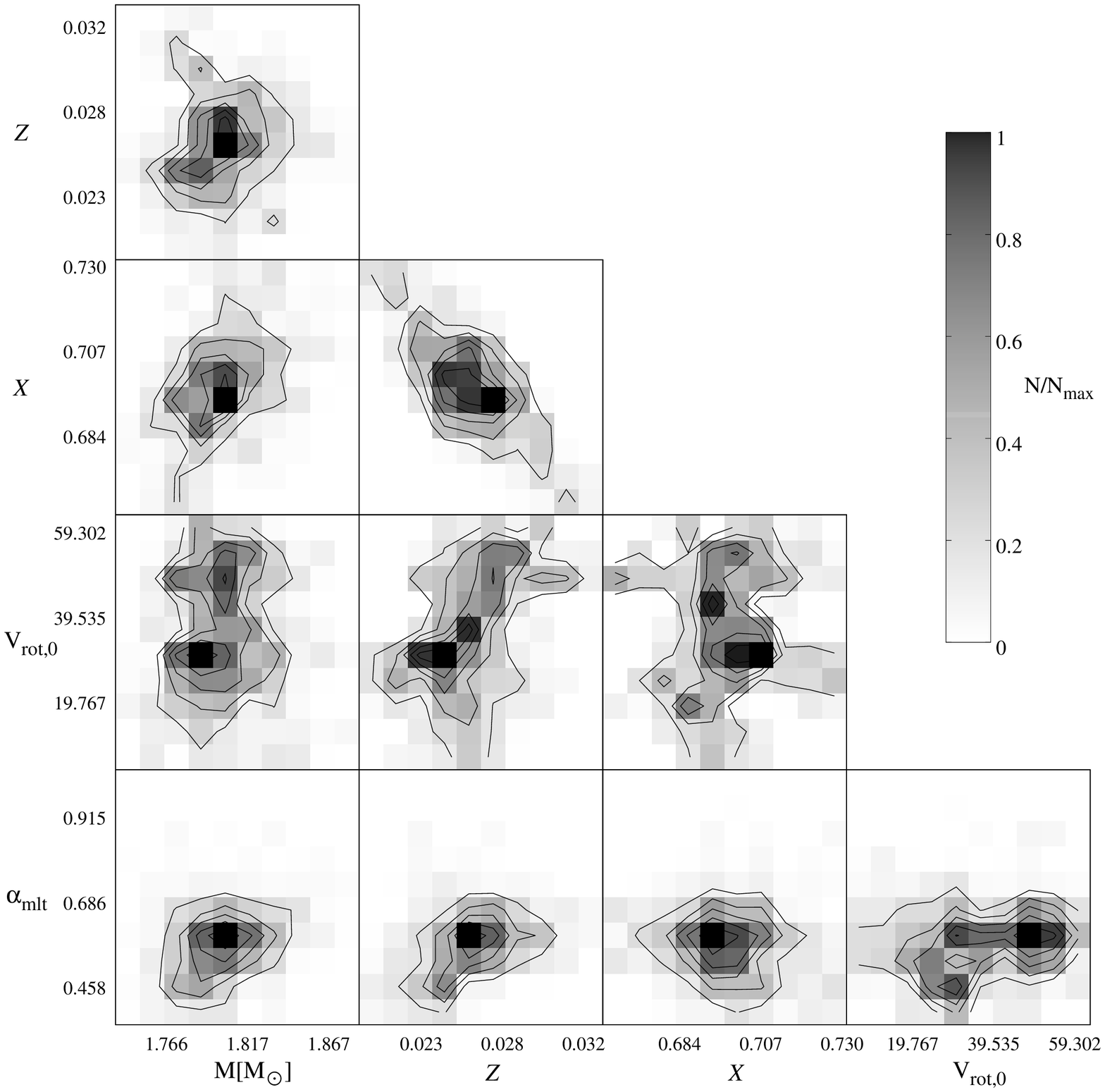}\par
		\includegraphics[width=88mm,clip]{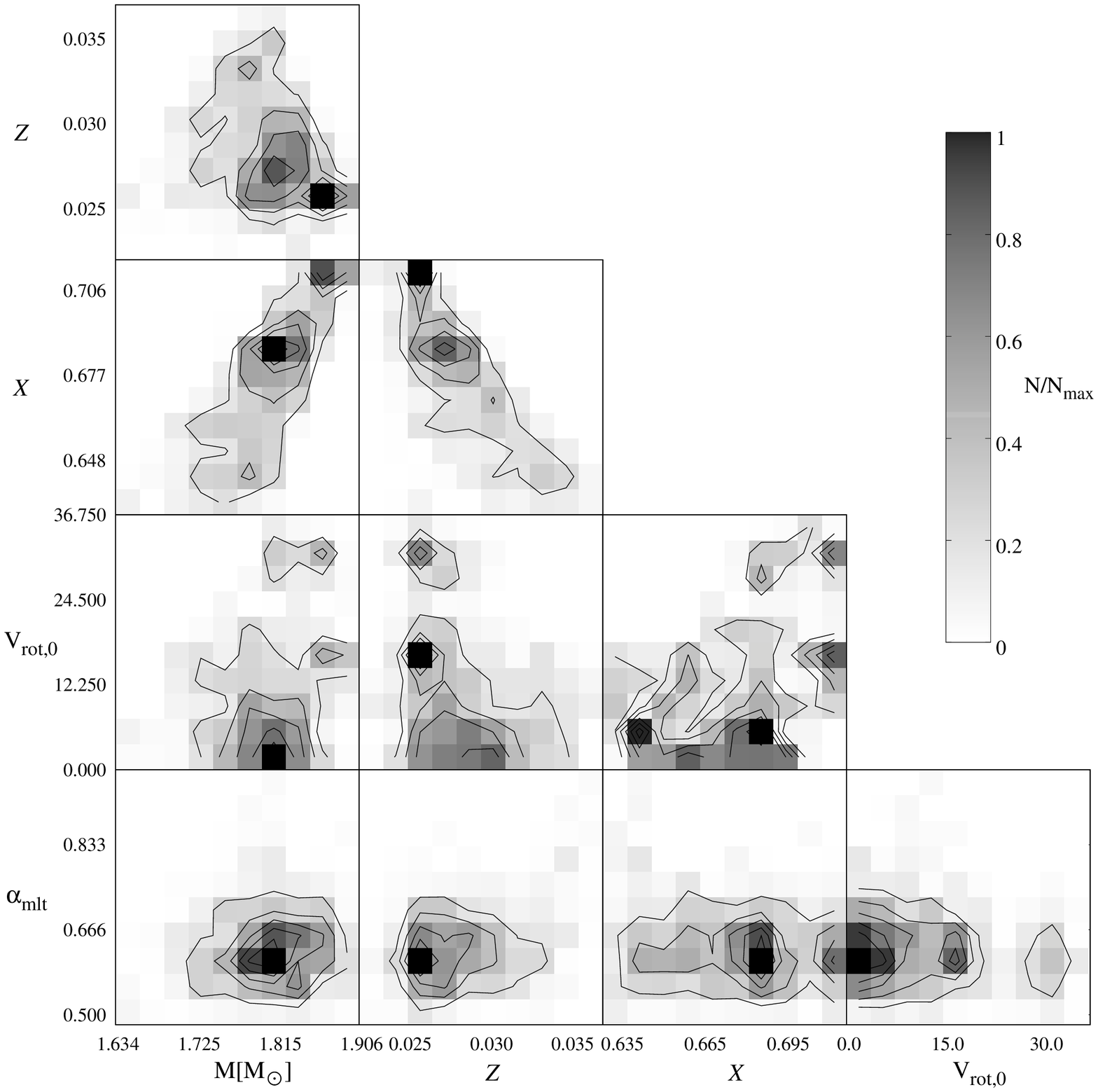}\par
	\end{multicols}
	\begin{multicols}{2}
		\includegraphics[width=88mm,clip]{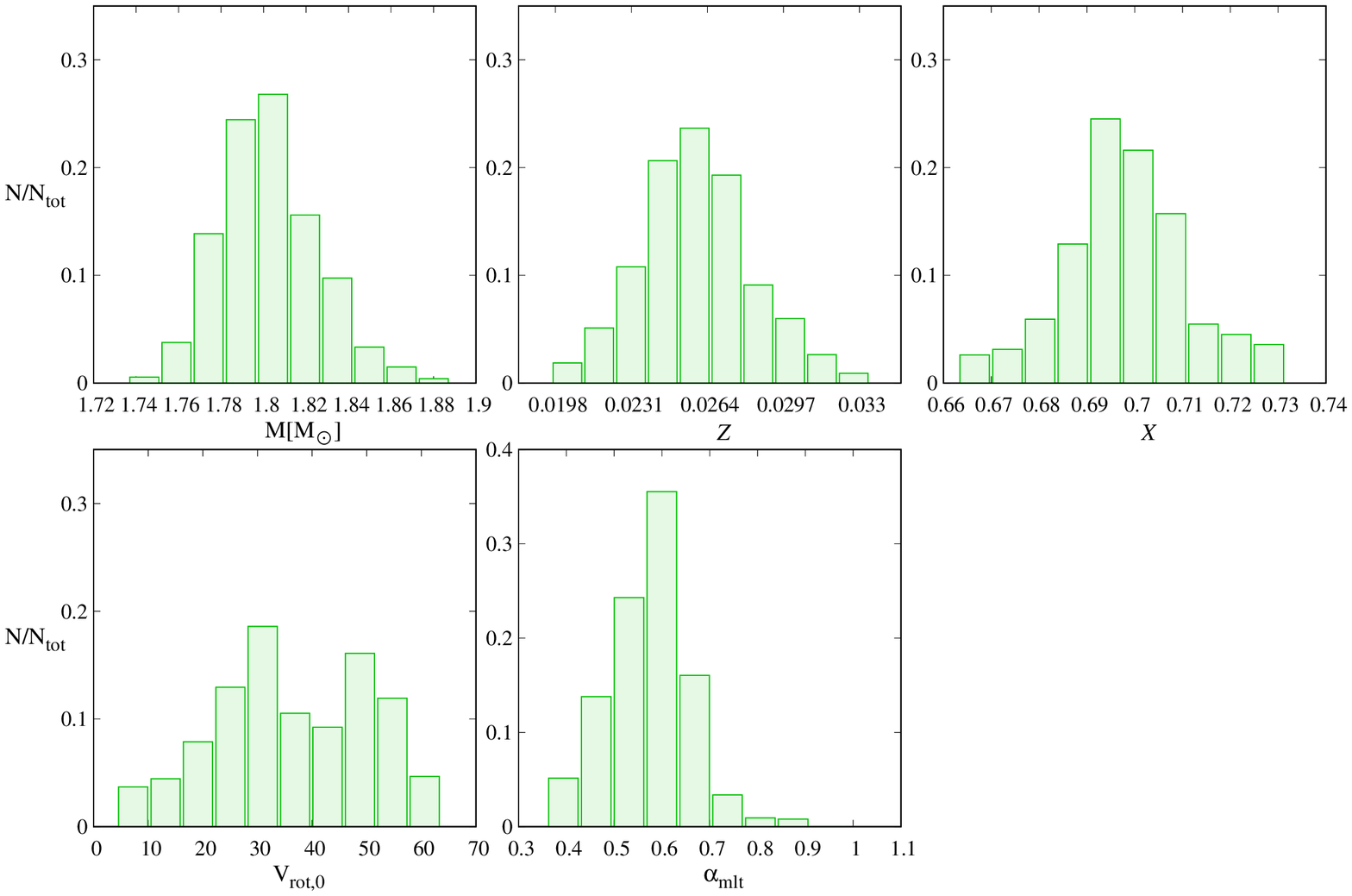}\par
		\includegraphics[width=88mm,clip]{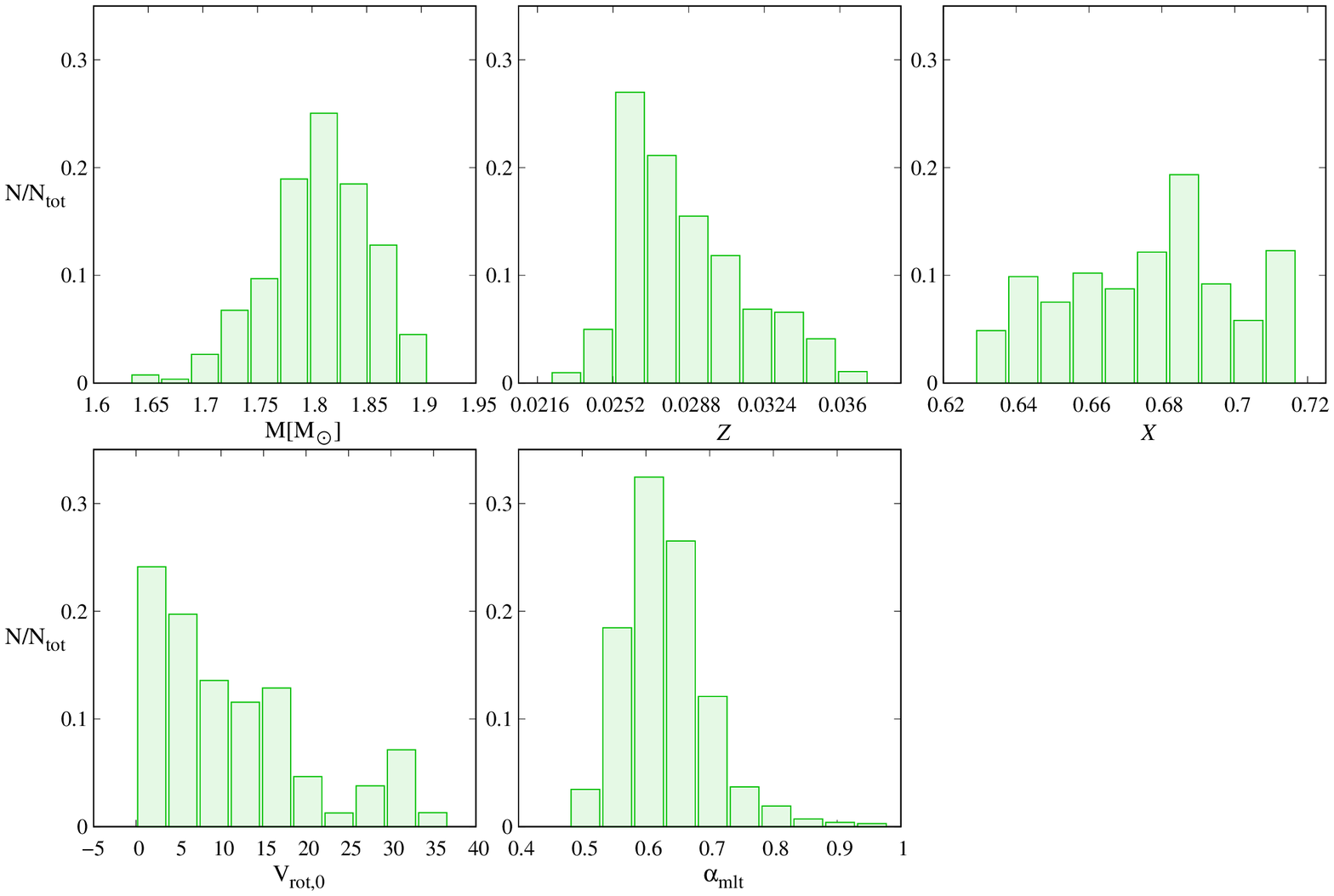}\par
	\end{multicols}	
	\caption{The same as in Fig.\ref{fig:mcs:cp1} but we included also the empirical parameter $f$ for the dominant mode
	 in our simulations. The NEMO model atmospheres were used with the microturbulent velocity $\xi_t=4$ km\,s$^{-1}$.}
	\label{fig:mcs:cp2}
\end{figure*}

\begin{figure}
	\includegraphics[angle=0, width=0.95\columnwidth]{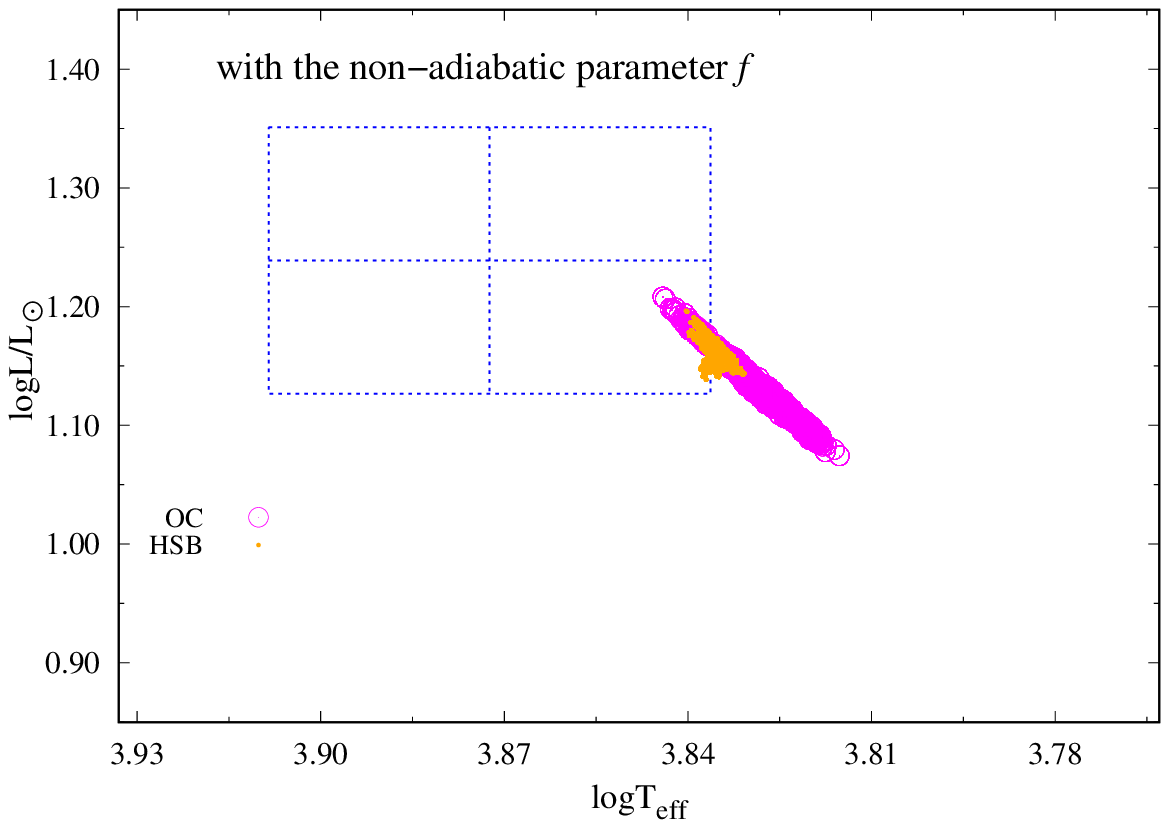}
	\caption{The same as in Fig.\,\ref{fig:mcs:HR1} but for the seismic models obtained from the Bayesian analysis including the parameter $f$ for the dominant mode. The NEMO model atmospheres were used with $\xi_t=4~\kms$.}
	\label{fig:mcs:HR3}
\end{figure}

The results of the simulations for $\xi_t=4~\kms$ are presented on the corner plots in Fig.\,\ref{fig:mcs:cp2}. As one can see, now the values of $\alpha_{\rm MLT}$ concentrates in a narrow range both for the OC and HSB models. Our simulations indicate rather small  convective efficiency with $\alpha_{\rm MLT}=0.58(8)$ for the OC seismic models
and $\alpha_{\rm MLT}=0.63(7)$ for the HSB seismic models.
The estimated mass of BP Peg are very similar
for the two evolutionary phases and amounts to $M=1.80(4)~{\rm M}_{\sun}$. The initial hydrogen abundance is $X_0=0.70(1)$
if BP Peg is in the OC phase or $X_0=0.67(2)$ in the HSB phase, whereas the  metallicity is $Z=0.026(2)$
in the OC phase or $Z=0.029(3)$ in the HSB phase.
The seismic  models from these simulations are also marked in the HR diagram shown in Fig.\,\ref{fig:mcs:HR3}.
As before all models are located near the bottom right corner of the error box.
The expected values of other parameters with the errors are given in Table\,6.  In general, all parameters are now better constrained
than in the case of simulations without the parameter $f$.  Moreover, one can also see that most parameters  are better determined
if the microturbulent velocity is $\xi_t=4~\kms$ or $\xi_t=8~\kms$. It is caused by the fact that the empirical values of $f$ for $\xi_t=2~\kms$ have very large errors  if [m/H]$<+0.3$, as has been shown in Fig.\,11 for one model (Sect.\,5.3).

For the other values of  the microturbulent velocity we got the following ranges of the mixing length parameter: \\
--   $\alpha_{\rm MLT}= 0.71(36)$ for the OC models and  $\alpha_{\rm MLT}= 0.16(10)$ for the HSB models if $\xi_t=2~\kms$ \\
 --  $\alpha_{\rm MLT}= 0.97(16)$ for the OC models and  $\alpha_{\rm MLT}= 0.71(8)$ for the HSB models if $\xi_t=8~\kms$.
Thus, in all cases the mixing parameter is below 1.0.

The seismic models obtained with $\xi_t=2~\kms$ have the initial hydrogen abundance of $X_0=0.64(2)$ in the OC phase
and  $X_0=0.78(1)$ in the HSB phase. These values of $X_0$ are unusually low (the OC phase) or  high (the HSB phase).
Therefore, we  treat this solution as much less probable. In the case of the simulations with $\xi_t=8~\kms$ we got $X_0=0.75(2)$
in the OC phase and   $X_0=0.68(1)$ in the HSB phase. These abundances of $X_0$ can be rather still acceptable.
The metallicity in this case differs significantly between the OC and HSB models, i.e., $Z=0.021(2)$ (OC) vs. $Z=0.018(1)$  (HSB).
Because the metallicity correlates with the mass, likewise,  there is the large difference in the mass value,
i.e., $M =1.92(3) ~{\rm M}_{\sun}$  for the OC models vs. $M =1.62(4) ~{\rm M}_{\sun}$  for the HSB models.

The age of all obtained models is in the range of about (1.43,\,1.54 ) Gyr.  The rotational velocity is in the range of about
$V_{\rm rot}\in(12,~33)~\kms$ which is consistent with a crude estimation by \citet{Kim1989}.

\section{Conclusions and future plans}
BP Pegasi is the post-main sequence $\delta$ Scuti star in the phase of the overall contraction
or hydrogen-shell burning.  Its age is  about 1.5 Gyr.
Without any doubts it pulsates in the two radial modes, the fundamental and first overtone.
There is no indication of any other variability from the observational data collected so far, neither in the older $uvby$ photometry
nor in the ASAS-3 data.

Using the amplitudes and phases in the $uvby$ bands we performed the mode identification of the two frequencies of BP Peg
confirming independently of the period ratio that these are two radial modes. As a by-product we derived the empirical values of
the amplitude of the bolometric flux variation (the parameter $f$) and the intrinsic mode amplitude $\varepsilon$.

Then, we made seismic modelling with the three sources of the opacity data showing that only with the OPAL data
it is possible to obtain seismic models of BP Peg  within the error box on the HR diagram if the mixing length parameter is below 2.0.
All seismic models that fit into the observed error box are in the post-main sequence phase of evolution.
We demonstrated also that the value of $\alpha_{\rm MLT}$ can have a huge effect on the Petersen diagram
and in some cases causes a deep decrease of the frequency ratio $\nu_1/\nu_2$ near TAMS. How big this effect is depends on mass
and metallicity. The decrease of $\nu_1/\nu_2$ is a consequence of changes in the internal structure of a star
described by the mean opacity profile $\kappa(T,\rho)$.

In the next step, we made an extensive seismic modelling with the Baysian analysis based on the Monte Carlo simulations.
Firstly, we showed that the convective overshooting is inefficient and it is reasonable to assume $\alpha_{\rm ov}=0.0$.
From, a huge number of simulations (about 80 000) we constrained the model parameters, i.e., the mass $M$, initial hydrogen abundance $X_0$, metallicity $Z$, rotational velocity $V_{\rm rot}$.  Moreover, we obtained that the efficiency of envelope convection
is characterized by the mixing  length parameter of about  $\alpha_{\rm MLT}= 0.5 - 1.0$ depending on the adopted
microturbulent velocity that we estimated at $\xi_t=4$ or 8\,$\kms$.
Thus, the convective transport in the envelope of BP Peg is rather moderately efficient. This conclusions is similar to our previous
results for SX Phe \citep{JDD2020} or for the prototype  $\delta$ Scuti  \citep{JDD2021}.
Definitely, such studies have to be continued to collect such seismic results for more $\delta$ Sct stars. 
Then, perhaps it will be possible to draw more general conclusions and of greater statistical significance,
on convection in the envelopes of stars with masses between 1.5 and 2.5\,${\rm M}_{\sun}$.
Another result that requires careful study is the effect of opacities on seismic models.
Why are  the OPAL seismic models data so different from the OP and OPLIB seismic models when
the two radial-mode frequencies are fitted?
Why are the seismic OPAL models better in a sense described in the paper? This is the second double-mode $\delta$ Scuti star
for which such result has been obtained. In order to try to answer these questions in the near future, we plan to enlarge the sample of double-radial mode $\delta$ Sct stars studied in this way.

\section*{Acknowledgements}
The work was financially supported by the Polish NCN grant 2018/29/B/ST9/02803.
Calculations have been partly carried out using resources provided by Wroclaw Centre for Networking and Supercomputing (http://www.wcss.pl), grant No. 265.
This work has made use of data from the European Space Agency (ESA) mission
{\it Gaia} (\url{https://www.cosmos.esa.int/gaia}), processed by the {\it Gaia}
Data Processing and Analysis Consortium (DPAC,
\url{https://www.cosmos.esa.int/web/gaia/dpac/consortium}). Funding for the DPAC
has been provided by national institutions, in particular the institutions
participating in the {\it Gaia} Multilateral Agreement.

\section*{Data Availability}
The ASAS observations are available at the website of \url{http://www.astrouw.edu.pl/asas}.
Theoretical computations will be shared on reasonable request to the corresponding author.



\bibliographystyle{mnras}
\bibliography{JDD_biblio} 


\bsp	
\label{lastpage}
\end{document}